\documentclass[reprint,showpacs,preprintnumbers,aps,prd]{revtex4-1}
\usepackage[T1]{fontenc}
\usepackage{color}
\usepackage{array}
\usepackage{multirow}
\usepackage{amsmath}
\usepackage{graphicx}

\makeatletter

\providecommand{\tabularnewline}{\\}

\usepackage{array}
\usepackage{dcolumn}% Align table columns on decimal point
\usepackage{longtable}
\usepackage[colorlinks,linkcolor=red,anchorcolor=green,citecolor=blue,CJKbookmarks=True]{hyperref}
\usepackage{bm}% bold math

\usepackage[dvipsnames]{xcolor}

\makeatother

\begin{document}
\title{Study of the pseudoscalar glueball in $J/\psi$ radiative decays}
\author{Long-Cheng Gui $^{1,2,3}$}
\email{guilongcheng@hunnu.edu.cn}

\author{Jia-Mei Dong $^{1}$}

\author{Ying Chen $^{4,5}$}
\email{cheny@ihep.ac.cn}

\author{Yi-Bo Yang $^{6}$}
\email{ybyang@itp.ac.cn}

\affiliation{{\small{}{}$^{1}$Department of Physics , Hunan Normal University,
ChangSha, 410081 , China~}~~\\
 {\small{}{} $^{2}$Key Laboratory of Low-Dimensional Quantum Structures
and Quantum Control of Ministry of Education, Changsha 410081, China
~}~~\\
 {\small{}{} $^{3}$Synergetic Innovation Center for Quantum Effects
and Applications(SICQEA), Hunan Normal University, Changsha 410081,China~}~~\\
 {\small{}{} $^{4}$Institute of High Energy Physics, Chinese Academy
of Sciences, Beijing 100049, China~}~~\\
 {\small{}{} $^{5}$School of Physics, University of Chinese Academy
of Sciences, Beijing 100049, China}~~\\
 {\small{}{} $^{6}$Institute of Theoretical Physics, Chinese Academy
of Sciences, Beijing 100190, China~}}
\begin{abstract}
We aim to explore the production rate of the pseudoscalar glueball
in $J/\psi$ radiative decay by lattice QCD in quenched approximation.
The calculation is performed on three anisotropic lattices with the
spatial lattice spacing ranging from 0.222(2) fm to 0.110(1) fm. As
a calibration of some systematical uncertainties, we first extract
the $M1$ form factor $\hat{V}(0)$ of the process $J/\psi\to\gamma\eta_{c}$
and get the result $\hat{V}(0)=1.933(41)$ in the continuum limit,
which gives the partial width $\Gamma(J/\psi\to\gamma\eta_{c})=2.47(11)$
keV. These results are in agreement with that of previous lattice
studies. As for the pseudoscalar glueball $G_{0^{-+}}$, its mass
is derived to be $2.395(14)$ GeV, and the form factor $\hat{V(0)}$
of the process $J/\psi\to\gamma G_{0^{-+}}$ is determined to be $\hat{V}(0)=0.0246(43)$
after continuum extrapolation. Finally, the production rate of the
pseudoscalar glueball is predicted to be $2.31(90)\times10^{-4}$,
which is much smaller than that of conventional light $q\bar{q}$
$\eta$ states. After the subtraction of the phase space factor, the
couplings of $J/\psi X\gamma$ are similar where $X$ stands for $\eta$
states and the pseudoscalar glueball. Possibly, the $U_{A}(1)$ anomaly
plays an important role for the large couplings of gluons to the flavor singlet $\eta$ states in $J/\psi$ radiative decays.
\end{abstract}
\maketitle

\section{INTRODUCTION}

\label{sec1} Quantum chromodynamics (QCD) predicts the existence
of glueballs, namely, the bound states of gluons. Last several decades
witnessed the intensive and extensive investigations of glueballs
both in experiments and theoretical studies\citep{Morningstar1999,Chen2006,Hao:2005hu,Nussinov:2009tq,Richards2010,Gregory2012,PhysRevD.87.054036,Qiao:2014vva,Sun2017a,PhysRevD.97.014007,Klempt:2007cp,Ochs:2013gi}.
Experimentally, there are ten scalar mesons observed with approximately
degenerated masses around 1.5 GeV, among which $f_{0}(1370)$, $f_{0}(1500)$,
and $f_{0}(1710)$ are the three isoscalars. According to the quark
model, if these states can be sorted into the $q\bar{q}$ $SU(3)$
nonet, then the surplus one isoscalar hints the existence of an additional
degree of freedom, possibly a glueball states, which can be either
one of the isoscalars mentioned above, or mixes with the conventional
$q\bar{q}$ states. A similar consideration applies to the pseudoscalar
channel: the three isoscalar pseudoscalar mesons $\eta(1295)$, $\eta(1405)$,
and $\eta(1495)$ also motivate the conjecture of the existence of
a pseudoscalar glueball in this mass range. Actually, there are many
phenomenological studies assigning $\eta(1405)$ to be the most likely
candidate for the pseudoscalar glueball due to its large production
fraction in the $J/\psi$ radiative decays. However, this assignment
has tensions with the prediction of the pseudoscalar glueball mass
from lattice QCD.

Lattice QCD is the {\it ab initio} nonperturbative approach for solving
QCD and plays a key role in the investigation of the low energy strong
interaction phenomena. In the glueball sector, lattice QCD in the
quenched approximation predicts that the masses of the lowest lying
scalar, tensor and pseudoscalar glueballs are roughly $1.5-1.7$ GeV,
$2.2-2.4$ GeV, and $2.6$ GeV, respectively\citep{Chen2006,Morningstar1999}.
Recent lattice calculations with dynamical quarks seemingly support
these predictions and do not observe large unquenched effects\citep{Richards2010,Gregory2012}.
As far as the pseudoscalar glueball is concerned, its predicted mass,
say, around 2.6 GeV, is much higher than that of $\eta(1405)$. This
discrepancy cannot be easily mediated by considering the glueball-meson
mixing in the presence of dynamical quarks. It is interesting to notice
that some phenomenological studies advocate $\eta(1405)$ and $\eta(1495)$
be the same state which appears differently in different final states
due to some dynamical mechanism\citep{Qin2017}. If this is the case,
then there is no redundant pseudoscalar meson mass region and subsequently
no need for an additional degree of freedom such as a glueball in
$1.5$ GeV mass region. As such, one may wish to search for the pseudoscalar
glueball in the energy range beyond 2 GeV according to the lattice
predictions.

$J/\psi$ radiative decays are regarded as an important hunting ground
for glueballs, owing to its the gluon-rich environment and cleaner
background. Apart from their masses, the production rates of glueballs
in $J/\psi$ radiative serve as additional key criteria for the identification
of glueballs, if they can be derived reliably from the theoretical
calculation. Also some quenched lattice QCD efforts have been made
to calculate these production rates of the scalar and tensor glueballs
\citep{Gui2013,Yang2013a}. Since the pure gauge glueballs are well
defined hadron states in the quenched approximation, the electromagnetic
form factors of $J/\psi$ radiatively decaying into glueballs can
be extracted directly by calculating the matrix elements of the electromagnetic
current between $J/\psi$ and glueballs. With these form factors,
the branch fraction of $J/\psi$ radiative decaying into the pure
gauge scalar glueball is predicted to be $3.8(9)\times10^{-3}$. It
is interesting to notice that the sum of the observed branching fraction
of the processes $J/\psi\rightarrow\gamma f_{0}(1710)\rightarrow\gamma+anything$
gives a value of roughly $2.0\times10^{-3}$ which is very close to
the above predicted value for the pure gauge glueball, while that
of $J/\psi\rightarrow\gamma f_{0}(1500)$ is an order of magnitude
smaller.

Recently, the BESIII collaboration has performed the partial wave
analysis to the process $J/\psi\rightarrow\gamma\phi\phi$ and observed
a new resonance $X(2500)$ with the resonance parameters $M_{X}=2470_{-19-23}^{+15+101}$
and $\Gamma_{X}=230_{-35-33}^{+64+56}$\citep{Ablikim2016a}. In the
process $J/\psi\rightarrow\gamma\eta'\pi^{+}\pi^{-}$, BESIII observed
a complicated structure in the $\eta'\pi^{+}\pi^{-}$ invariant mass
spectrum and reported two new resonances $X(2120)$ and $X(2370)$,
whose quantum number are likely $J^{PC}=0^{-+}$\citep{Ablikim2011}.
These new resonances lie in the mass range of the pseudoscalar glueball
predicted by lattice QCD and are worthy of further experimental investigations.
However, in order to unravel the nature of these states, more theoretical
inputs are desired.

In this work, we calculate the production rate of the pseudoscalar
glueball from lattice QCD by adopting the similar strategy in the
scalar and tensor cases. Even though the dynamical lattice QCD
simulation is dominant nowadays, the study of glueballs is still challenging
since it requires much higher statistics comparing to the usual hadron.
Therefore, we would like to perform an exploratory investigation by generating
the quenched gauge configurations with large statistics using the
anisotropic lattices. In order to calibrate part of uncertainties,
we first calculate the partial width of the process $J/\psi\rightarrow\gamma\eta_{c}$
and compare with the results of previous lattice calculations and
experiments. We admit that glueballs may have strong mixing with conventional
mesons in the real world, and \textcolor{red}{{} }the quenched effects
on our result of glueball production rates cannot be reliably estimated
in the present stage.

This work is organized as follows: Section II gives an introduction to the
formalism for calculating the radiative transition width of $J/\psi$ from lattice QCD.
Section III presents the calculation details and results including
the parameters of the lattice, the relevant spectrum, and
transition form-factors. We give the conclusion and some discussions
in Section IV.

\section{FORMALISM}

\global\long\def\amp{\mathcal{M}_{\lambda_{J/\psi},\lambda_{\gamma}}}%
\global\long\def\phoe{\epsilon_{\mu}^{*}(\vec{k},\lambda_{\gamma})}%
\global\long\def\staG{X(\vec{p}_{f})}%
\global\long\def\staJ{J/\psi(\vec{p}_{i},\lambda_{J/\psi})}%
\global\long\def\vecq{\vec{q}}%
Generally speaking, in the rest frame of $J/\psi$, the partial decay width of $J/\psi$ radiatively
decaying into a pseudoscalar meson $X$ can be calculated through
the following formula,
\begin{equation}
\Gamma_{J/\psi\to\gamma X}=\frac{1}{24\pi}\frac{|\vec{k}|}{M_{J/\psi}^{2}}\sum_{\lambda_{J/\psi},\lambda_{\gamma}}|\mathcal{M}_{\lambda_{J/\psi},\lambda_{\gamma}}(\vec{p}_{i},\vec{p}_{f})|^{2},
\end{equation}
where $\vec{p}_{i}$ and $\vec{p}_{f}$ are the momenta of $J/\psi$
and $X$, respectively, $\vec{k}=\vec{p}_i-\vec{p}_{f}=-\vec{p}_f$ is the
decaying momentum of the emitted photon,
$\lambda_{J/\psi}$ and $\lambda_{\gamma}$ stand for the different
polarizations of $J/\psi$ and the photon, and $\mathcal{M}_{\lambda_{J/\psi},\lambda_{\gamma}}$
is the on-shell transition amplitude. The magnitude of $\vec{k}$ can
be defined through the masses of $J/\psi$ (denoted by $M_{J/\psi}$)
and $X$ ( denoted by $M_{X}$ ), say, $|\vec{k}|=\frac{M_{J/\psi}^{2}-M_{X}^{2}}{2M_{J/\psi}}$.
The transition amplitude contains all the dynamics of the decay, and
can be expressed to the lowest order of QED as
\begin{equation}
\amp=\epsilon_{\mu}^{*}(\vec{k},\lambda_{\gamma})\langle X(\vec{p}_{f})|J_{{\rm em}}^{\mu}|J/\psi(\vec{p}_{i},\lambda_{J/\psi})\rangle,
\end{equation}
where $\phoe$ is the polarization vector of photon and $J_{{\rm em}}^{\mu}=\sum\limits _{f}e_{f}\bar{\psi}_{f}\gamma^{\mu}\psi_{f}$
is the relevant electromagnetic vector current with the summation
over all possible quark flavors. If the sea quark contributions through
disconnected diagrams can be neglected, one can use $J_{{\rm em}}^{\mu}=e_{c}\bar{c}\gamma_{\mu}c$
as an approximation for the electromagnetic decays of charmonia. Usually,
the matrix elements $\langle X(\vec{p}_{f})|J_{\rm em}^{\mu}(0)|J/\psi(\vec{p}_{i},\lambda_{J/\psi})\rangle$
can be expressed in terms of form factor through the multipole decomposition.
For the pseudoscalar $X$, the explicit expression is
\begin{eqnarray}
\langle X(\vec{p}_{f})|J_{{\rm em}}^{\mu}(0)|\staJ\rangle & = & M(Q^{2})\epsilon^{\mu\nu\rho\sigma}p_{i,\nu}p_{f,\rho}\nonumber \\
 & \times & \epsilon_{\sigma}(\vec{p}_{i},\lambda_{J/\psi})\label{eq:multipoleformfactor}
\end{eqnarray}
where $Q^{2}\equiv-(p_{i}-p_{f})^{2}$ and $M(Q^{2})$ is the multipole
form factor, which is sometimes also expressed in terms of a dimensionless
form factor $V(Q^{2})$ as $M(Q^{2})=\frac{2V(Q^{2})}{m_{X}+m_{J/\psi}}$.
Thereby the partial decay width can be written as
\begin{eqnarray}
\Gamma_{J/\psi\to\gamma X} & = & \frac{1}{12\pi}|\vec{k}|^{3}\left|M(0)\right|^{2}\nonumber \\
 & \equiv & \frac{1}{3\pi}\frac{|\vec{k}|^{3}}{(m_{J/\psi}+m_{X})^{2}}\left|V(0)\right|^{2}.
\end{eqnarray}
It is clearly seen that if the matrix elements in Eq.~(\ref{eq:multipoleformfactor})
are known, the form factor $M(Q^{2})$ (or equivalently $V(Q^{2})$)
can be derived to give the decay width directly. Actually, this goal
can be achieved in lattice QCD study by calculating the relevant three-point
correlation functions
\begin{eqnarray}
\Gamma_{J/\psi\to\gamma X}^{(3),\mu i}(\vec{p}_{i},t_{i}=0;&&\vec{p}_{f},t_{f};\vecq,t)  =  \sum_{\vec{x},\vec{y},\vec{z}}e^{i\vec{p}_{i}\cdot\vec{x}}e^{i\vec{q}\cdot\vec{y}}e^{-i\vec{p}_{f}\cdot\vec{z}}\nonumber \\
 &  & \langle\mathcal{O}_{X}(\vec{z},t_{f})J_{{\rm em}}^{\mu}(\vec{y},t)\mathcal{O}_{J/\psi,i}^{\dagger}(\vec{x},t_{i})\rangle,\nonumber \\
\end{eqnarray}
where $\mathcal{O}_{X}$ and $\mathcal{O}_{J/\psi,i}$ are the interpolating
field operators for $X$ and $J/\psi$, respectively. After the intermediate
state insertion, the three-point function $\Gamma_{J/\psi\to\gamma X}^{(3),\mu i}$
is parametrized as
\begin{widetext}
\begin{equation}
\begin{array}{ccc}
\Gamma_{J/\psi\to\gamma X}^{(3),\mu i}(\vec{p}_{i},t_{i}=0;\vec{p}_{f},t_{f};\vecq,t) & = & \sum\limits _{m,n,\lambda_{n}}\frac{e^{-E_{m}t_{f}}e^{-(E_{n}-E_{m})t}}{4E_{m}E_{n}}\langle\Omega\mid\mathcal{O}_{f}(0)\mid m,\vec{p}_{f}\rangle\\
 & \times & \langle m,\vec{p}_{f}\mid J_{{\rm em}}^{\mu}(0)\mid n,\vec{p}_{i},\lambda_{n}\rangle\langle n,\vec{p}_{i},\lambda_{n}\mid\mathcal{O}_{J/\psi,i}^{\dagger}(0)\mid\Omega\rangle\\
 & \xrightarrow{t_{f}\gg t\gg0} & \frac{e^{-E_{X}t_{f}}e^{-(E_{J/\psi}-E_{X})t}}{4E_{J/\psi}E_{X}}Z_{i}^{*(J/\psi)}Z^{(X)}\langle\staG\mid J_{em}^{\mu}(0)\mid\staJ\rangle
\end{array}\label{eq:threepf}
\end{equation}
\end{widetext}

where $E_{X}$ and $E_{J/\psi}$ are the energies of $X$ and $J/\psi$,
respectively, $Z_{i}^{(J/\psi)}=\langle J/\psi(\vec{p}_{f},\lambda_{J/\psi})|\mathcal{O}_{J/\psi,i}(0)|\Omega\rangle$,
and $Z^{(X)}=\langle X(\vec{p}_{f})|\mathcal{O}_{X}(0)|\Omega\rangle$,
which can be extracted from the relevant two-point functions
\begin{eqnarray}
\Gamma_{ij}^{(2)}(\vec{p},t) & = & \sum e^{-i\vec{p}\cdot\vec{x}}\langle\Omega\mid\mathcal{O}_{i}(\vec{x},t)\mathcal{O}_{j}^{\dagger}(0,0)\mid\Omega\rangle\nonumber \\
 & \xrightarrow{t\to\infty} & \frac{Z_{i}Z_{j}^{*}}{2E(\vec{p})}e^{-E(\vec{p})t}.\label{eq:twopf}
\end{eqnarray}
Practically, one can carry out a joint fit to the two-point functions
and the three-point function to extract the desired matrix elements
in Eq.~(\ref{eq:multipoleformfactor}), from which the multipole
form factors can be derived at different $Q^{2}$. % & =\sum_{n,\lambda}\frac{1}{2E}\left\langle \Omega\mid\mathcal{O}_{i}(t)\mid %n(\vec{p},\lambda)\rangle\langle %n(\vec{p},\lambda)\mid\mathcal{O}_{j}^{\dagger}(0)\mid\Omega\right\rangle \\
%\begin{equation}
%\Gamma_{J/\psi\to\gamma %G_{ps}}^{(3)}(\vec{p}_{J/\psi,}\vec{p}_{G_{ps}},\vecq,t_{i},t,t_{f})=\sum_{\vec{x},\vec{y},\vec{z}}e^{-i\vec{p}_{J/\psi}\cdot\vec{x}}e^{i\vecq\cdot\vec{y}}e^{-i\vec{p}_{G_{p%s}}\cdot\vec{z}}\times\langle\mathcal{O}_{G_{ps}}(\vec{z},t_{f})J_{em}^{\mu}(\vec{y},t)\mathcal{O}_{J/\psi}^{\dagger}(\vec{x},t_{i})\rangle
%\end{equation}
%where $\mathcal{O}_{G_{ps}},\mathcal{O}_{J/\psi}$ are the interpolating
%operators which have the same quantum number with pseudoscalar glueball
%and $J/\psi$, respectively. We can get the hadron matrix elements
%by inserting complete sets of states at both sides of the electromagnetic
%vector current:

%where $Z^{(J/\psi)},Z^{(G_{ps})}$ are the overlap matrix elements
%for $J/\psi$ and pseudoscalar interpolating operators like $Z^{(n)}=\langle\Omega|\mathcal{O}|n\rangle$.
%The energy $E_{J/\psi},E_{G_{ps}}$and overlap matrix elements $Z^{(J/\psi)},Z^{(G_{ps})}$
%can be derived from relevant two-point functions

\section{NUMERICAL DETAILS}

As addressed in Sec.~\ref{sec1}, we perform the calculation in the
quenched approximation. Since $J/\psi$, $\eta_{c}$, and the pseudoscalar
glueball are heavy particles, in order to obtain good signals with
high resolutions in the temporal direction, we generate the gauge
configurations on anisotropic lattices with the temporal lattice much
finer than the spatial lattice, say, $\xi=a_{s}/a_{t}\gg1$, where
$a_{s}$ and $a_{t}$ are the spatial and temporal lattice spacings,
respectively. In practice, we choose $\xi=5$. The gauge action we
use is the tadpole improved gauge action~\citep{Morningstar1997}
whose discretization error is expected to be $O(a_{s}^{4},\alpha_{s}a_{s}^{2})$.
Three gauge ensembles with large statistics are generated at different
lattice spacings for the continuum extrapolation, and the relevant
ensemble parameters are listed in Table\ref{tab:lattice}, where
the lattice spacings $a_{s}$ are determined from $r_{0}^{-1}=410(20)\,{\rm MeV}$
by calculating the static potential. For fermions, we use the tadpole
improved clover action for anisotropic lattices~\citep{Liu2002b}.
The parameters in the action are tuned carefully by requiring that
the physical dispersion relations of vector and pseudoscalar mesons
are correctly reproduced at each bare quark mass~\citep{Zhang2001}.
The bare charm quark masses for the two lattices are set by the physical
mass of $J/\psi$, $m_{J/\psi}=3.097\,{\rm GeV}$.

%In our calculation, we use tadpole improved gauge action{[}{]} and
%tadpole improved clover fermion action{[}{]} in the quenched approximation
%lattice QCD. Anisotropic lattices be adopted to perform simulation
%which give small $m_{c}a_{t}$ and fine temporal direction resolution
%to hadron correlation function. Our simulation perform on two anisotropic
%lattices with $\xi=5$, namely $L^{3}\times T=12^{3}\times144$ and
%$16^{3}\times160$. The relevant input parameters are listed in Table?,
%where $a_{s}$ values are determined from $r_{0}^{-1}=410(20)$MeV.
%We generate 20000 configurations for $12^{3}\times144$ lattice and
%10000 configurations for $16^{3}\times160$ lattice. The parameters
%in the action are tuned carefully by requiring that the physical dispersion
%relation of vector and pseudoscalar are correctly reproduced at each
%bare quark mass{[}{]}. The bare quark mass be set to reproduce $m_{J/\psi}=3.097$
%GeV.
%\par\end{flushleft}
%%%%%%%%%%%%%%%%%%%%%%%%%%%%%%%%%%%%%%%%%%%%%%%%%%%%%%%%%%%%%%%%%%%%%%%%%%%%%%%%%%%%%%%%%%%%%
\begin{table}
\caption{\label{tab:lattice} Relevant input parameters in this work. The
spatial lattice spacing $a_{s}$ is determined from $r_{0}^{-1}=410(20)\,{\rm MeV}$
by calculating the static potential.}

\begin{ruledtabular}
\begin{tabular}{cccccc}
$\beta$  & $\xi$  & $a_{s}$(fm)  & $La_{s}$(fm)  & $L^{3}\times T$  & $N_{conf.}$\tabularnewline
\hline
2.4  & 5  & 0.222(2)  & 1.78  & $8^{3}\times96$  & 20000\tabularnewline
2.8  & 5  & 0.138(1)  & 1.66  & $12^{3}\times144$  & 20000\tabularnewline
3.0  & 5  & 0.110(1)  & 1.76  & $16^{3}\times160$  & 10000\tabularnewline
\end{tabular}
\end{ruledtabular}

\end{table}
%%%%%%%%%%%%%%%%%%%%%%%%%%%%%%%%%%%%%%%%%%%%%%%%%%%%%%%%%%%%%%%%%%%%%%%%%%%%%%%%%%%%%%%%%%%%%
%%%%%%%%%%%%%%%%%%%%%%%%%%%%%%%%%%%%%%%%%%%%%%%%%%%%%%%%%%%%%%%%%%%%%%%%%%%%%%%%%%%%%%%%%%%%%%%%
\begin{figure}[t]
\includegraphics[scale=0.4]{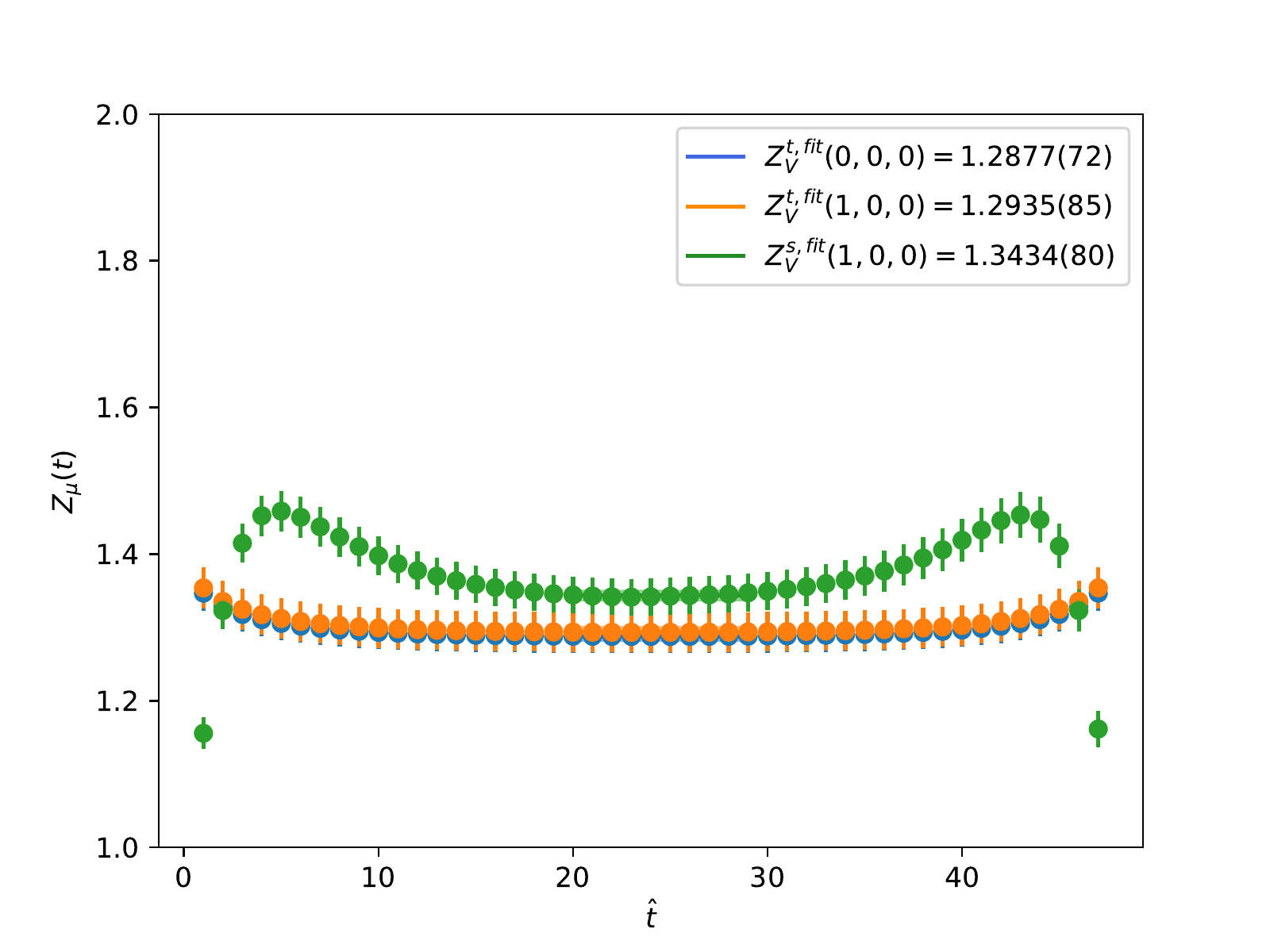}\\
 \includegraphics[scale=0.4]{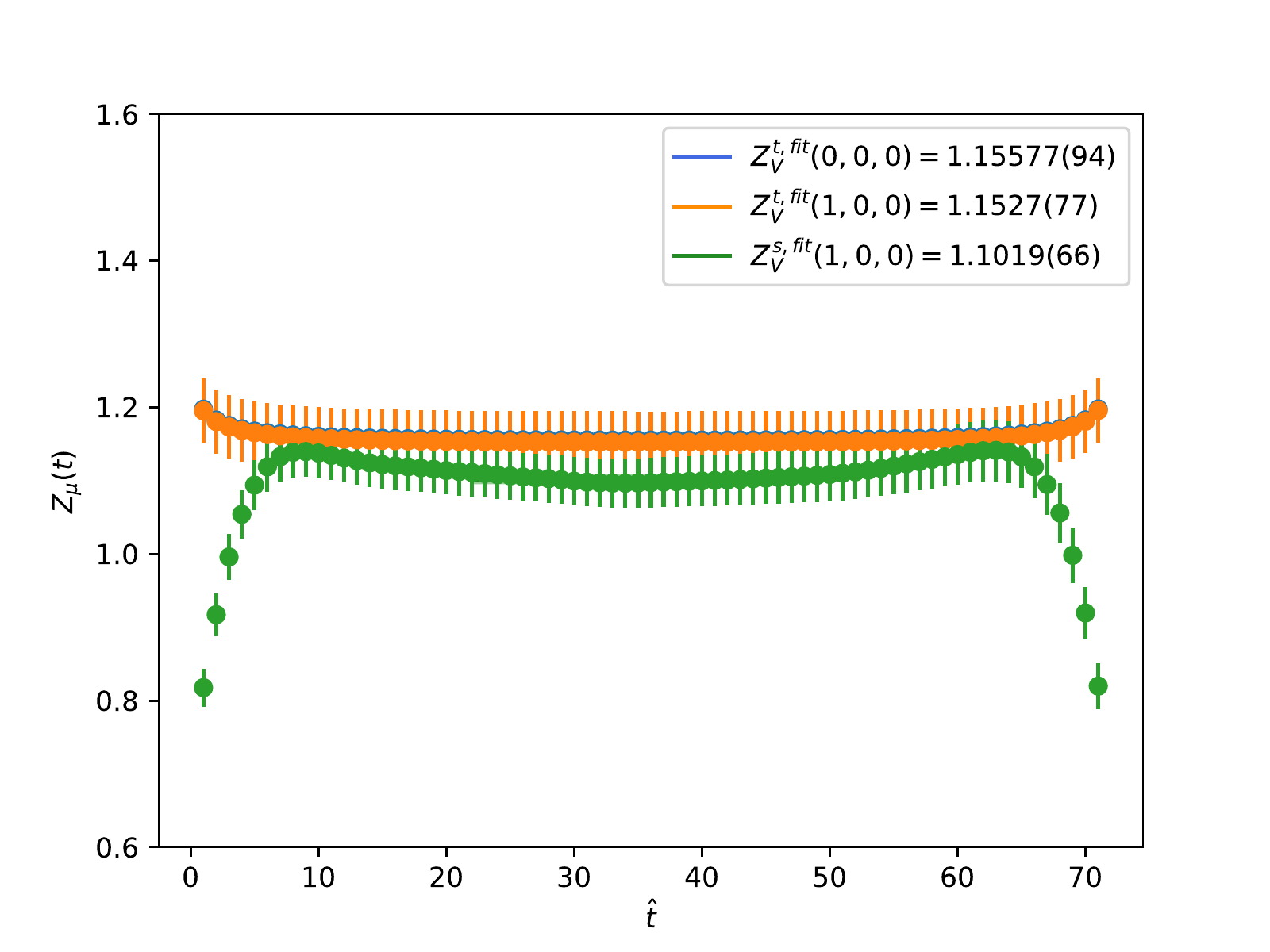}\\
 \includegraphics[scale=0.4]{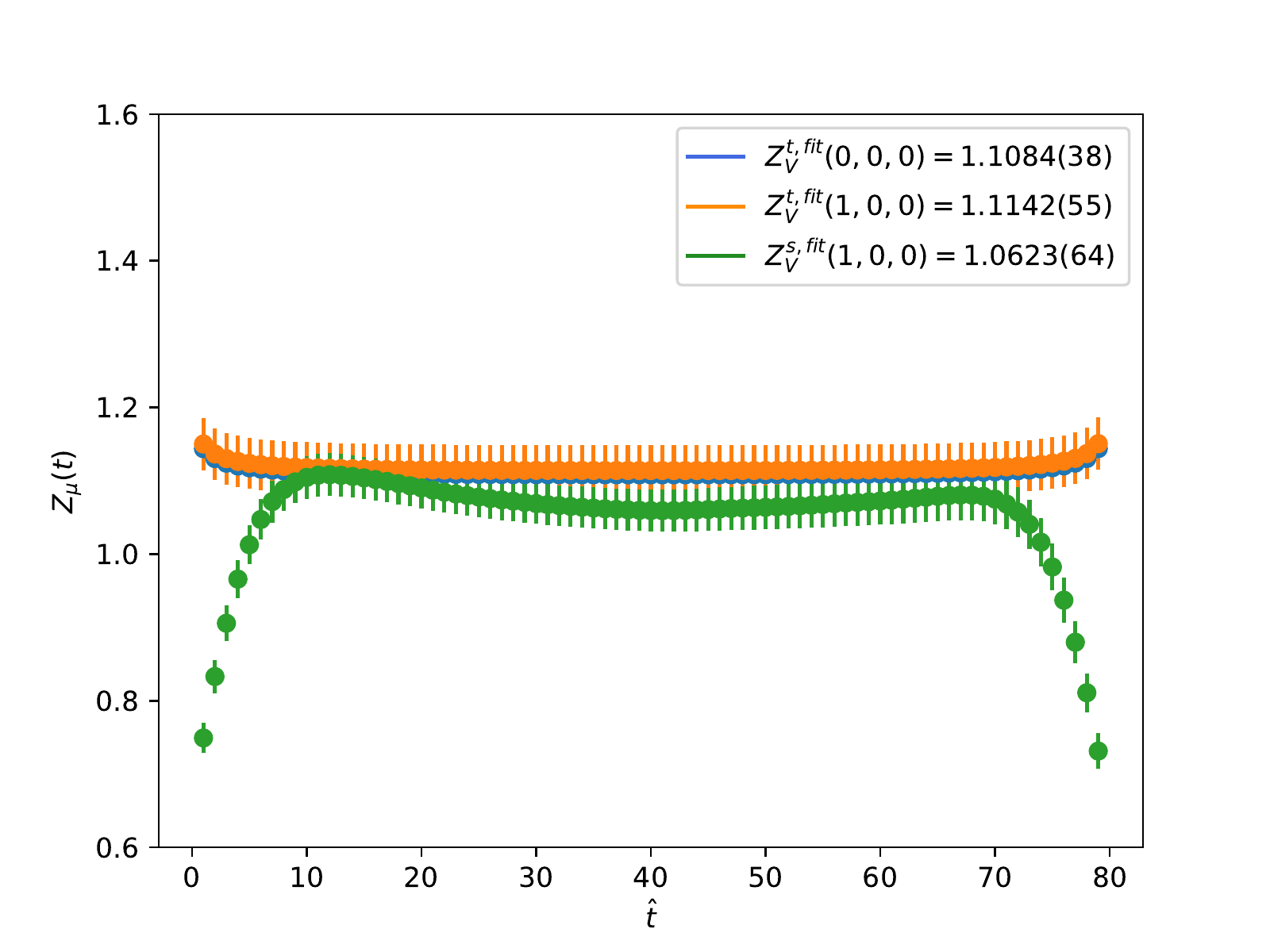}\\
 \caption{\label{fig:renorm_cur} The renormalization factor $Z_{V}^{(s)}$ and $Z_{V}^{(t)}$
with respect to $t$ for the three lattices. The panels from top to bottom correspond
to $\beta=2.4,2.8,3.0$, respectively}
\end{figure}
%%%%%%%%%%%%%%%%%%%%%%%%%%%%%%%%%%%%%%%%%%%%%%%%%%%%%%%%%%%%%%%%%%%%%%%%%%%%%%%%%%%%%%%%%%%%%%%%
In the quenched approximation, since there are no sea quarks, the
electromagnetic current contributing to the radiative transitions
of charmonia involves only the charm quark, say, $J_{{\rm em}}^\mu(x)=e_{c}J^{\mu}(x)$
with $J^{\mu}(x)=\bar{c}\gamma_{\mu}c(x)$, which is the one we adopt
in this study. It is a conserved vector current and need not be renormalized
in the continuum. However, on a finite lattice, it is not conserved
anymore due to the lattice artifacts and receives a multiplicative
renormalization factor $Z_{V}(a_{s})$. Following the scheme proposed
by Ref.~\citep{Dudek2005a}, $Z_{V}(a_{s})$ is extracted using the
ratio of the $\eta_{c}$ two-point function and the related three-point
function evaluated at $Q^{2}=0$,
\begin{equation}
Z_{V}^{(\mu)}(t)=\frac{p^{\mu}}{E(\vec{p})}\frac{\frac{1}{2}\Gamma_{\eta_{c}}^{(2)}(\vec{p};t_{f}=\frac{n_{t}}{2})}{\Gamma^{(3),\mu}(\vec{p},\vec{p},\frac{n_{t}}{2},t)},
\end{equation}
where the factor $1/2$ accounts for the effect of the temporal periodic
boundary condition, and the superscript $\mu$ of $Z_{V}(a_{s})$
is used to differentiate the temporal component from the spatial ones,
since they are not necessarily the same on the anisotropic lattices. Figure~\ref{fig:renorm_cur} plots $Z_{V}^{(\mu)}(t)$ with
respect to $t$ for the three lattices. $Z_{V}$'s are extracted from the plateaus and the values
are listed in Table~\ref{renorm}. Obviously, the renormalization constant $Z_V^{(s)}(a)$ of the spatial components of the vector current deviate from that of the temporal component, $Z_V^{(t)}(a)$ by a few percents. This deviation is understandable since 
the space-time interchange symmetry is broken on anisotropic lattices, apart from 
the imperfect tuning of the bare speed of light in the fermion action. 
In practice, we perform the calculation in the rest frame of the final pseudoscalar state. 
In this frame, according to Eq.~(\ref{eq:multipoleformfactor}), the matrix element of the temporal component of the vector current is zero due to the appearance of the totally antisymmetric tensor $\epsilon^{\mu\nu\rho\sigma}$, and we extract the form factor $M(Q^2)$ only from the corresponding matrix elements of the spatial components of the vector current. So we only need the renormalization factor $Z_{V}^{(s)}$.
%%%%%%%%%%%%%%%%%%%%%%%%%%%%%%%%%%%%%%%%%%%%%%%%%%%%%%%%%%%%%%%%%%%%%%%%%%%%%%%%%%%%%%%%%%%%%%%%
\begin{table}
\caption{The renormalization constants $Z_{V}^{(s)}$ and $Z_{V}^{(t)}$ of
the spatial and temporal components of the vector current for $\beta=2.4$,
$\beta=2.8$ and $\beta=3.0$ lattices. Two momentum modes, $(0,0,0)$
and $(1,0,0)$, are used for the derivation.}
\label{renorm}
\begin{ruledtabular}
\begin{tabular}{cccc}
$\beta$  & $Z_{V}^{(t)}(0,0,0)$  & $Z_{V}^{(t)}(1,0,0)$  & $Z_{V}^{(s)}(1,0,0)$ \tabularnewline
\hline
$2.4$  & $1.288(5)$  & $1.299(11)$  & $1.388(15)$ \tabularnewline
$2.8$  & $1.155(3)$  & $1.159(3)$  & $1.110(7)$ \tabularnewline
$3.0$  & $1.106(4)$  & $1.114(6)$  & $1.062(6)$\tabularnewline
\end{tabular}
\end{ruledtabular}

\end{table}
%%%%%%%%%%%%%%%%%%%%%%%%%%%%%%%%%%%%%%%%%%%%%%%%%%%%%%%%%%%%%%%%%%%%%%%%%%%%%%%%%%%%%%%%%%%%%%%%

%\textcolor{red}{In this study, we adopt the current operator as $\bar{\psi}\gamma_{\mu}\psi$
%which is conserved in the continuum theory but not conserved on the
%lattice. So we need to extract additional renormalization factor $Z_{V}$
%on each lattice by the ratio of two-point function and three-point
%function evaluated at $Q^{2}=0$.}

%\textcolor{red}{
%\begin{equation}
%\begin{aligned}Z_{V}^{\mu}(x) & %=\frac{p^{\mu}}{E(\vec{p})}\frac{\sum_{i}\frac{1}{2}\Gamma_{\mathcal{O}_{i}\mathcal{O}_{i}}^{(2)%}(\vec{p},t_{f}=T/2)}{\sum_{i}\Gamma_{\mathcal{O}_{i}\gamma_{\mu}\mathcal{O}_{i}}^{(3)}(\vec{p}_%{f}=\vec{p_{i}}=\vec{p};t_{f}=T/2,t)}\end{aligned}
%\end{equation}
%The factor $\frac{1}{2}$ accounts for the effect of the temporal
%periodic boundary condition. The $\mathcal{O}_{i}$ denote the corresponding
%particle and $\eta_{c}$ be choiced in this calculation as there is
%no significant dependence upon the particle used \cite{Dudek2005a}.
%the renormalisation of the spatial and temporal components of the
%vector current have a slight difference due that we are working on
%anisotropic lattice. Figure {[}{]} show the $Z_{V}(t)$ dependency
%on t for three lattices. We using constant fit $Z_{V}$ on plateaus
%and the results list in Table . The values of $Z_{V}$ are consistent
%with our previous results on two smaller lattice . In this work, the
%renormalisation factor of the spatial components of the vector current
%$Z_{V}^{(s)}$ are used in all calculation. }

\subsection{$J/\psi\to\gamma\eta_{c}$ transition}

\begin{figure}
\includegraphics[width=0.4\textwidth]{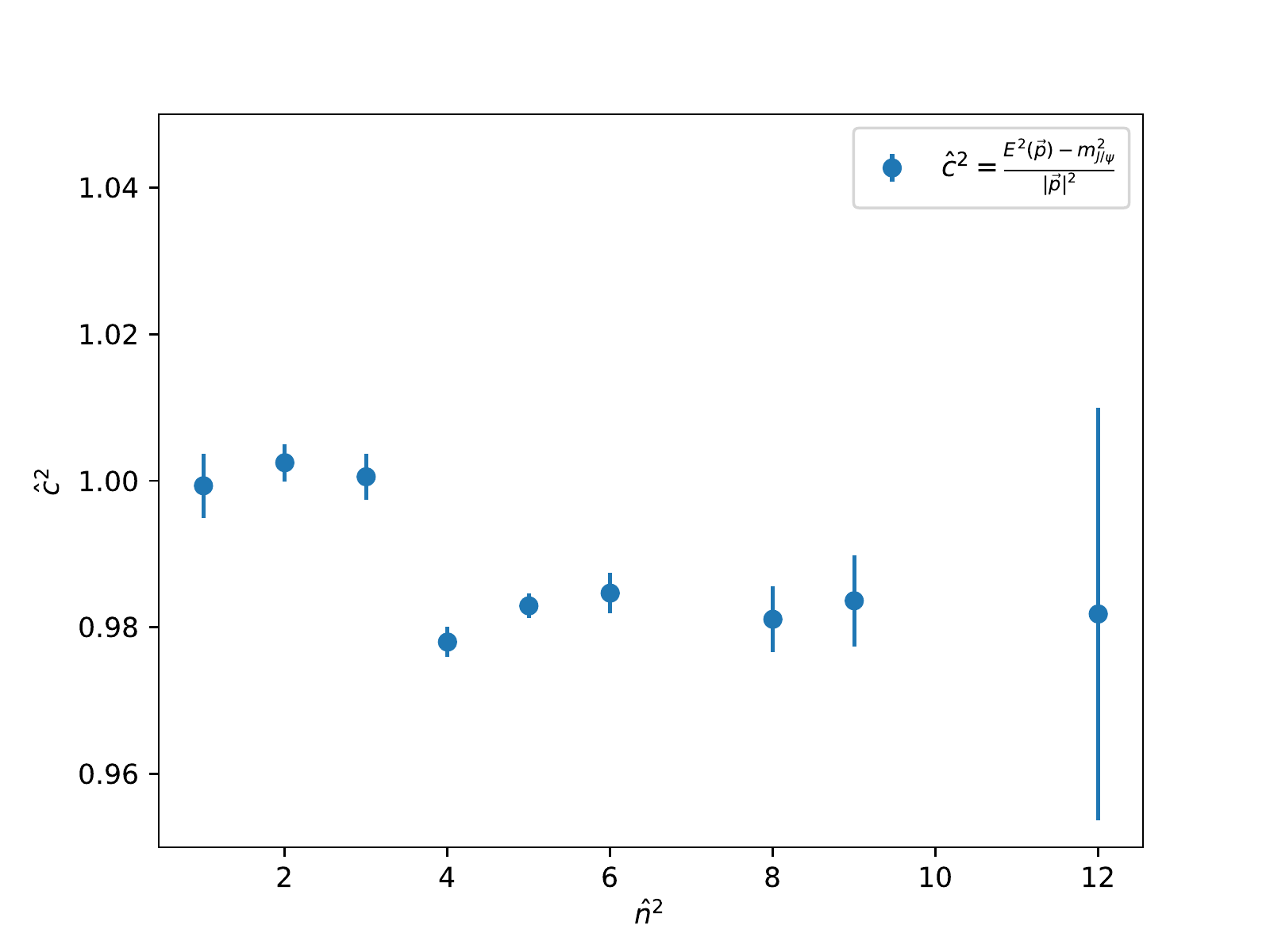}

\caption{The squared speed of light $\hat{c}^{2}$ versus the momentum modes $\hat{n}$ ranging from $(0,0,0)$ to $(2,2,2)$ on the $\beta=3.0$ lattice. Each data point shows the averaged value of $\hat{c}^2$ over the momentum modes $\hat{n}$ with the same $|\hat{n}|^2$ by assuming the 
    approximate rotational symmetry. The largest deviation of $\hat{c}^{2}$ from one is less than 4\%.}
\label{fig:vlight}
\end{figure}

There have been quite a few lattice studies on the decay process $J/\psi\rightarrow\gamma\eta_{c}$.
We would like to carry the similar calculation and make a comparison
with previous studies as well as the experimental value, which serve
as a calibration to some discretization uncertainties of our lattice
setup. We work in rest frame of the pseudoscalar (such as $\eta_{c}$)
with $J/\psi$ moving with a definite momentum $\hat{p}_{i}=2\pi\hat{n}/La_{s}$,
where $\hat{n}$ ranges from $(0,0,0)$ to $(2,2,2)$. %%%%%%%%%%%%%%%%%%%%%%%%%%%%%%%%%%%%%%%%%%%%%%%%%%%%%%%%%%%%%%%%%%%%%%%%%%%%%%%%%%%%%%%%%%%%%%%
\begin{figure}[t]
\includegraphics[scale=0.45]{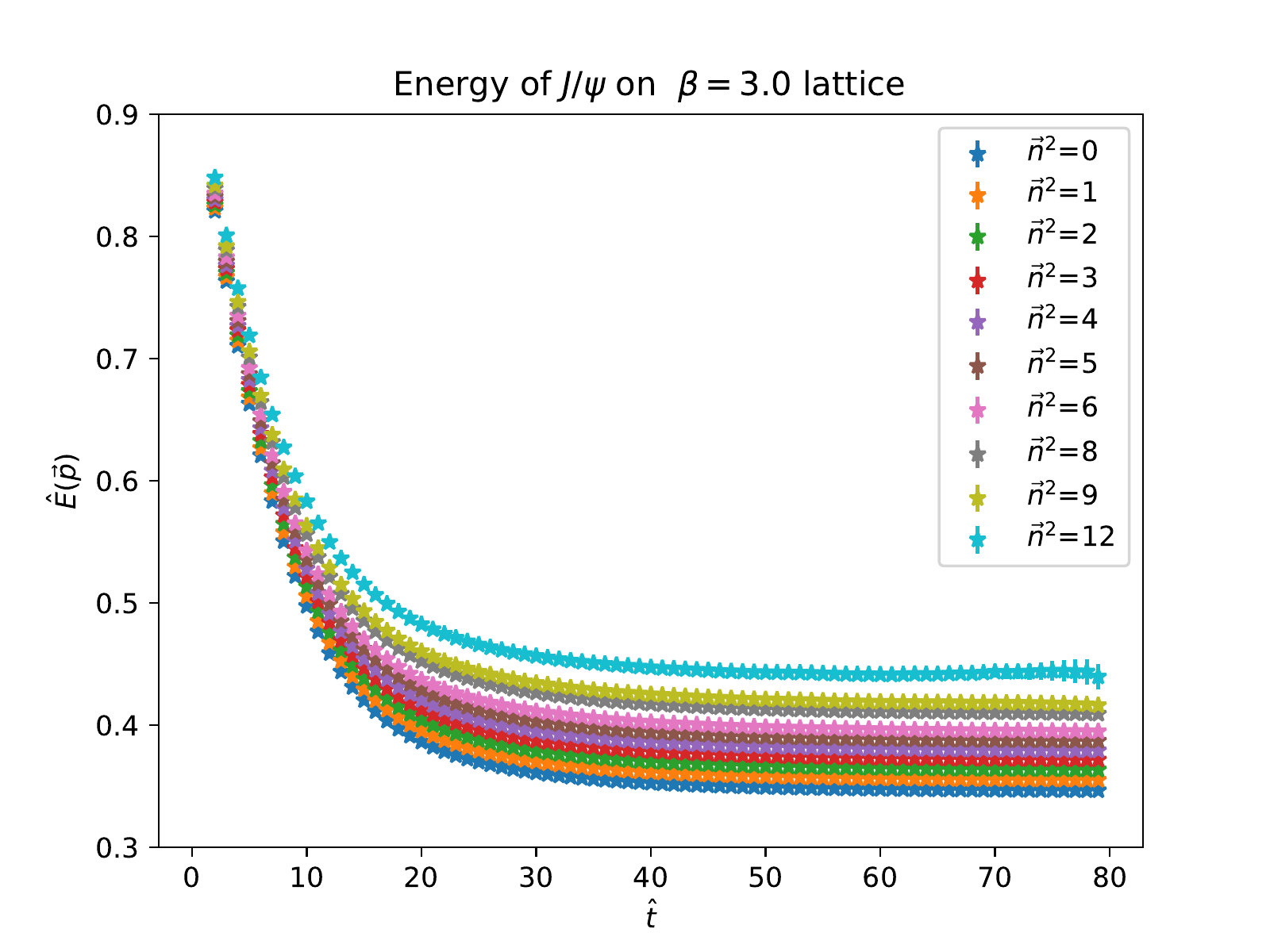}\\
 \caption{\label{fig:effectmass} Effective energy plateaus $E(\vec{p})$ of
$J/\psi$ for $\beta=3.0$. $E(\vec{p})$'s are averaged over the
momentum modes with the same $|\vec{p}|^{2}$. It is seen that signal-to-noise
ratios are good for the momentum modes we are using.}
\end{figure}
%%%%%%%%%%%%%%%%%%%%%%%%%%%%%%%%%%%%%%%%%%%%%%%%%%%%%%%%%%%%%%%%%%%%%%%%%%%%%%%%%%%%%%%%%%%%%%%
%%%%%%%%%%%%%%%%%%%%%%%%%%%%%%%%%%%%%%%%%%%%%%%%%%%%%%%%%%%%%%%%%%%%%%%%%%%%%%%%%%%%%%%%%%%%%%%%%
%\begin{figure}[t]
%\includegraphics[scale=0.35]{pic/beta24_vlight}	
%	\includegraphics[scale=0.35]{pic/beta28_vlight}
%	\includegraphics[scale=0.35]{pic/beta30_vlight}
%	\caption{\label{fig:vlight}Squared speed of light extracted from $J/\psi$ dispersion relation:
%		$c^{2}\equiv\frac{E^{2}(\vec{p})-m^{2}}{\mathbf{p}^{2}}$}
%\end{figure}
%%%%%%%%%%%%%%%%%%%%%%%%%%%%%%%%%%%%%%%%%%%%%%%%%%%%%%%%%%%%%%%%%%%%%%%%%%%%%%%%%%%%%%%%%%%%%%%%%
As mentioned above, both the three-point functions and two-point functions
are required in order to extract the desired hadronic matrix elements
of the current $J^{\mu}_{{\rm em}}$. We choose the quark bilinear
operators $\mathcal{O}_{i}(x)=(\bar{c}\Gamma_{i}c)(x)$ for $\eta_{c}$
($\Gamma_{i}=\gamma_{5}$) and $J/\psi$ ($\Gamma_{i}=\gamma_{i},i=1,2,3$),
such that the two-point function with momentum $\vec{p}$ can be calculated
through
\begin{equation}
\Gamma_{ij}^{(2)}(\vec{p},t)=-\sum\limits _{\vec{x}}e^{-i\vec{p}\cdot\vec{x}}\text{Tr}\langle S^{\dagger}(\vec{x},t;\vec{0},0)\Gamma_{i}S(\vec{x},t;\vec{0},0)\Gamma_{j}\rangle
\end{equation}
where $S(\vec{x},t,\vec{0},0)$ is the point-source propagator of
the charm quark. 
The effective energy plateaus $E(\vec{p})$ of $J/\psi$ are illustrated
in Fig.~\ref{fig:effectmass} for $\beta=3.0$, where $E(\vec{p})$'s are averaged over the
momenta with the same $|\vec{p}|^{2}$. We check
the dispersion relation of $J/\psi$ by calculating the squared speed of light $\hat{c}^2$
\begin{equation}
\hat{c}^{2}=\frac{E^{2}(\vec{p})-m_{J/\psi}^{2}}{|\vec{p}|^{2}}.
\end{equation}
It is found that the largest deviation
of $\hat{c}^{2}$ from one is less than 4\% on all
the three lattices. For illustration, we plot $\hat{c}^2$ with respect to 
different momentum modes on the $\beta=3.0$ lattice in Fig.~\ref{fig:vlight}, where the data points 
are the averaged values over the momentum modes $\hat{n}$ with the same $|\hat{n}|^2$ by assuming the 
approximate rotational symmetry. 

\begin{figure*}[t]
\includegraphics[width=0.3\textwidth]{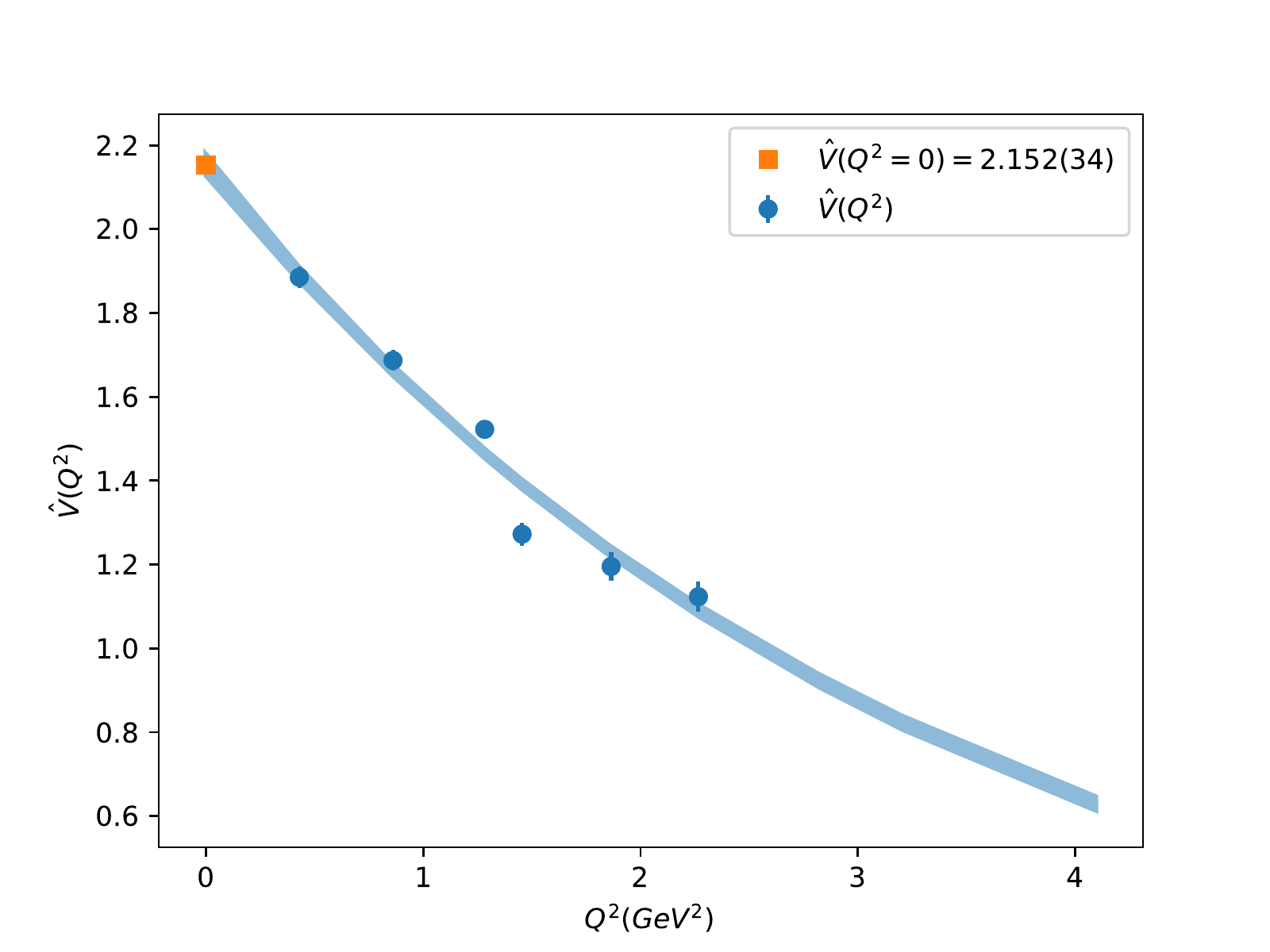}
\includegraphics[width=0.3\textwidth]{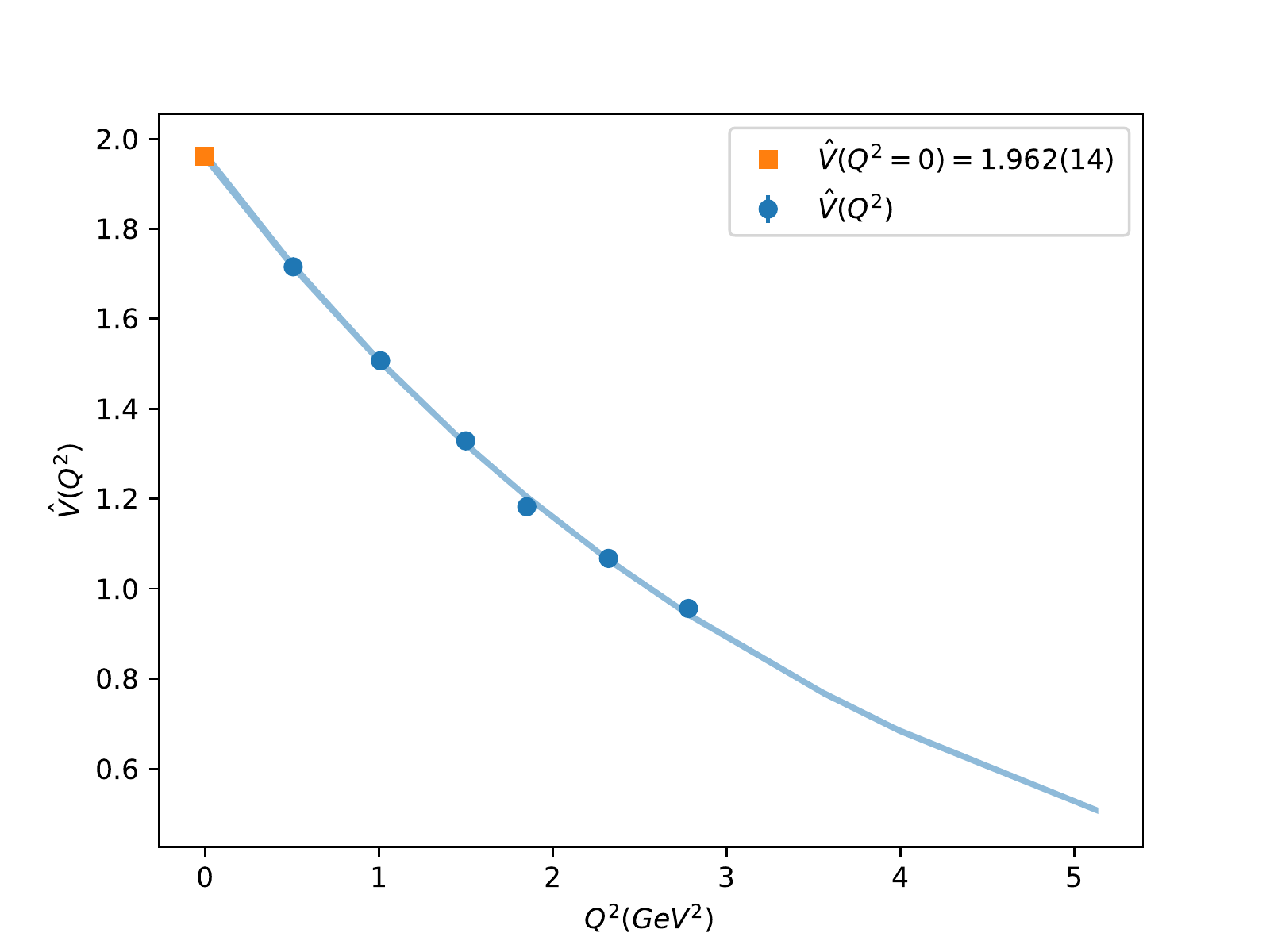}
\includegraphics[width=0.3\textwidth]{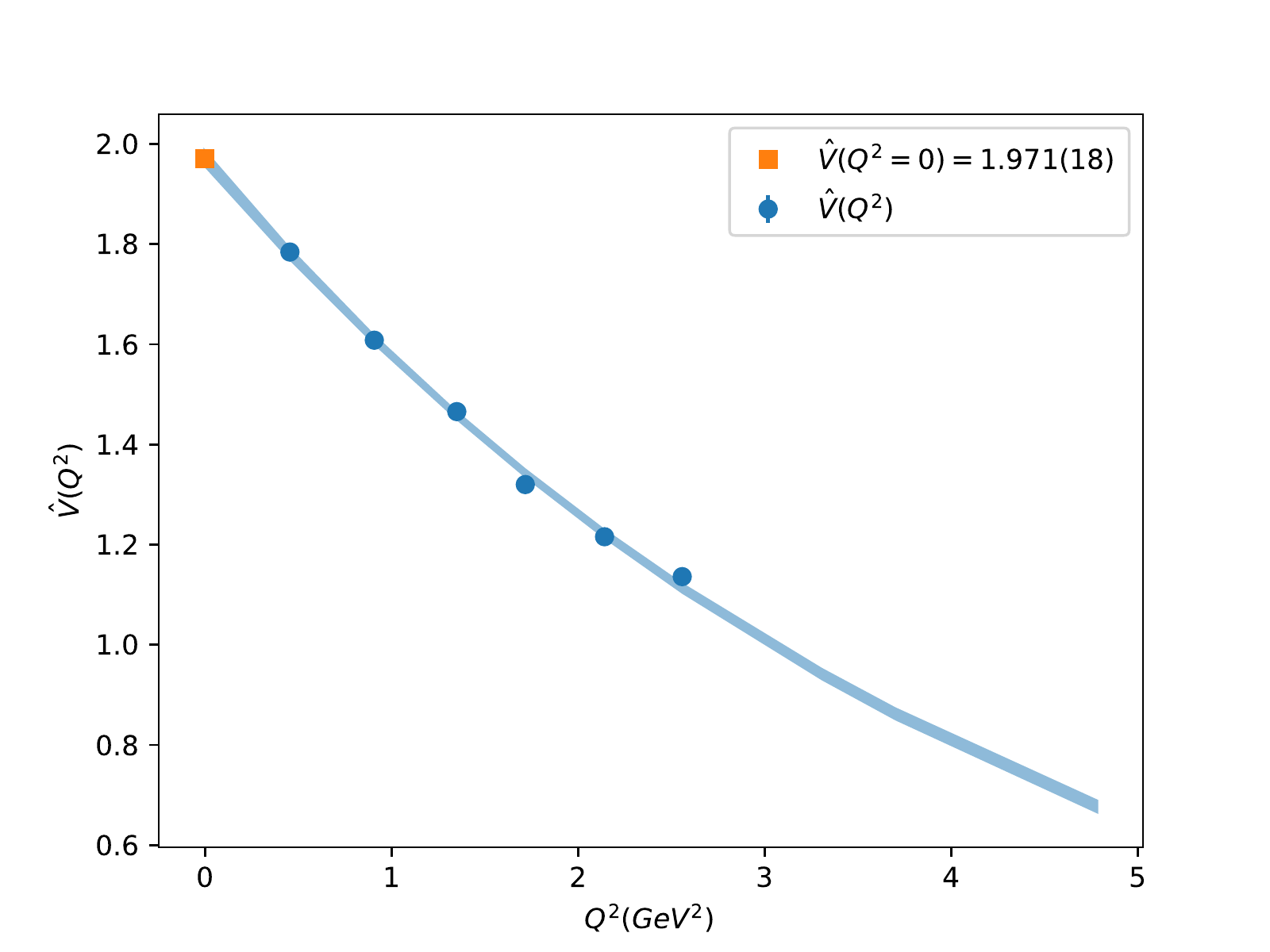}
\caption{\label{fig:jpsitogammaetac}The form factor $\hat{V}(Q^{2})$ in Eq.\ (\ref{eq:multipoleformfactor})
of $J/\psi$ to $\gamma\eta_{c}$ where $Q^{2}=-(P_{J/\psi}-P_{\eta_{c}})^{2}$.
Exponential functions in Eq.\ (\ref{eq:jpsitoetacfitf}) are adopted
for extrapolation. The panels from left to right corresponds to $\beta=2.4,2.8,3.0$
lattices, respectively.}
\end{figure*}

The three-point functions $\Gamma_{J/\psi\to\gamma\eta_{c}}^{(3),\mu i}$
contributed by the connected diagrams (disconnected diagrams are neglected)
are calculated through the expression
\begin{eqnarray}
&& \Gamma_{J/\psi\to\gamma\eta_{c}}^{(3),\mu i}(\vec{p}_{i},t_{i}=0;\vec{p}_{f},t_{f};\vecq,t) \nonumber \\ 
&& \quad =  \sum_{\vec{x},\vec{y}}e^{-i\vec{p}_f\cdot\vec{x}+i\vec{q}\cdot\vec{y}}\langle\mathcal{O}_{\Gamma_f}(\vec{x},t_{f}) (\bar{c}\gamma_{\mu}c)(\vec{y},t)\mathcal{O}_{i}^{\dagger}(\vec{0},0)\rangle\nonumber \\
&& \quad =  -\sum_{\vec{y}}e^{i\vec{q} \cdot \vec{y}}{\rm tr}\langle H_{\Gamma_{f}}^{\dagger}(\vec{y},t;\vec{0},0;\vec{p}_{f},t_{f}) \gamma_{5}\gamma_{\mu}S(\vec{y},t;\vec{0},0)\gamma_{i}\rangle \nonumber \\
&&
\end{eqnarray}
where
\begin{equation}
H_{\Gamma_{f}}(y,0;t_{f};\vec{p}_f)\equiv\sum_{\vec{x}}e^{i\vec{p}_{f} \cdot \vec{x}}S(\vec{y},t;\vec{x},t_f)\gamma_{5}S(\vec{x},t_f;\vec{0},0)\gamma_{5}
\end{equation}
can be obtained by the sequential source technique \citep{Dudek2005a}.
In order to increase the statistics, we repeat the same calculations
$T$ times (where $T$ is the temporal lattice size) by setting a
point source on a different time slice each time. With the related
two-point functions calculated accordingly, a straightforward way
to extract the interesting matrix elements $\langle\eta_{c}(\vec{p}_{f},\lambda_{f})|J_{{\rm em}}^{\mu}(0)|J/\psi(\vec{p}_{i},\lambda_{i})\rangle$
is to fit the three-point function and two-point function simultaneously
according to Eq.~(\ref{eq:threepf}) and Eq.~(\ref{eq:twopf}) through
the jackknife analysis. To suppress the contribution of excited states,
we use this formula %\begin{widetext}
\begin{eqnarray}
R^{\mu i}(\vecq,t_f,t) & = & \Gamma^{(3),\mu i}(\vec{q},t_{f},t)\sqrt{\frac{2E_{i}\Gamma_{i}^{(2)}(\vec{p}_{i},t_{f}-t)}{\Gamma_{i}^{(2)}(\vec{p}_{i},t)\Gamma_{i}^{(2)}(\vec{p}_{i},t_{f})}}\nonumber \\
 & \times & \sqrt{\frac{2E_{f}\Gamma_{f}^{(2)}(\vec{p}_{f},t)}{\Gamma_{f}^{(2)}(\vec{p}_{i},t_{f}-t)\Gamma_{i}^{(2)}(\vec{p}_{i},t_{f})}}
\end{eqnarray}
%\end{widetext}
which gives flatter plateaus. In practice, the energies $E_{i,f}$
are derived from two-point functions in the joint fit of the two-point
and three-point functions. We can get the form factors by solving
the Eq.\ (\ref{eq:multipoleformfactor}). Based on the OZI rule,
we neglect the contribution from the quark annihilation diagrams and
only consider the contribution of connected diagrams. As such we compute
the form factor $\hat{V}(Q^{2})$ which is related to $V(Q^{2})$
by
\begin{equation}
V(Q^{2})=2\times\frac{2}{3}e\times\hat{V}(Q^{2}),
\end{equation}
where the factor $2$ comes from the insertion of the electromagnetic
current to both the quark and antiquark lines, $2/3e$ is the electric
charge of charm quark. In the expression above, the renormalization
constant of the spatial component of $J^{\mu}_{{\rm em}}$, say,
$Z_{V}^{(s)}$, has been implicitly incorporated into $\hat{V}(Q^{2})$.
The extracted $\hat{V}(Q^{2})$ on the three lattices we are using
are plotted in terms of $Q^{2}$ in Fig.~\ref{fig:jpsitogammaetac}.
The $\beta=2.4$ lattice is coarse ($L^3\times T=8^{3}\times96$), such that results 
has larger systematic errors. We try to fit with the data at the smallest three $Q^{2}$ and get
$\hat{V}(Q^{2})=2.045(36)$. We regard the difference of the fitted values
using different $Q^{2}$ range as systematic error. The errors on other two lattices 
are statistical errors.

%\begin{equation}
%\Gamma_{J/\psi\to\eta_{c}\gamma}=\frac{16}{27}\alpha\frac{|\vec{q}|^{3}}{(m_{\eta_{c}}+m_{J/\psi})^{2}}|\hat{%V}(0)|^{2}
%\end{equation}
%where $\alpha=e^{2}/4\pi$is the fine structure constant about equal
%to $1/134$ on the charmonium energy region.
%%%%%%%%%%%%%%%%%%%%%%%%%%%%%%%%%%%%%%%%%%%%%%%%%%%%%%%%%%%%%%%%%%%%%%%%%%%%%%%%%%%%%%%%%%%%%%%%

%%%%%%%%%%%%%%%%%%%%%%%%%%%%%%%%%%%%%%%%%%%%%%%%%%%%%%%%%%%%%%%%%%%%%%%%%%%%%%%%%%%%%%%%%%%%%%%%
%%%%%%%%%%%%%%%%%%%%%%%%%%%%%%%%%%%%%%%%%%%%%%%%%%%%%%%%%%%%%%%%%%%%%%%%%%%%%%%%%%%%%%%%%%%%%%%%
\begin{table}
\caption{The mass of $J/\psi$, $\eta_{c}$ and the form factor $\hat{V}(0)$
on three lattices. \label{tab:jpsitogammaetac}}

\begin{ruledtabular}
\begin{tabular}{cccc}
$\beta$  & $m_{J/\psi}$ (GeV)  & $m_{\eta_{c}}$ (GeV)  & $\hat{V}(0)$ \tabularnewline
\hline
2.4  & 3.097(1)  & 2.995(1)  & 2.152(34)(107) \tabularnewline
2.8  & 3.102(1)  & 3.007(2)  & 1.962(14) \tabularnewline
3.0  & 3.105(1)  & 2.995(1)  & 1.971(18) \tabularnewline
\hline
$\infty$  &  &  & 1.933(41) \tabularnewline
\end{tabular}
\end{ruledtabular}

\end{table}
%%%%%%%%%%%%%%%%%%%%%%%%%%%%%%%%%%%%%%%%%%%%%%%%%%%%%%%%%%%%%%%%%%%%%%%%%%%%%%%%%%%%%%%%%%%%%%%%

\begin{figure}[t]
\includegraphics[scale=0.45]{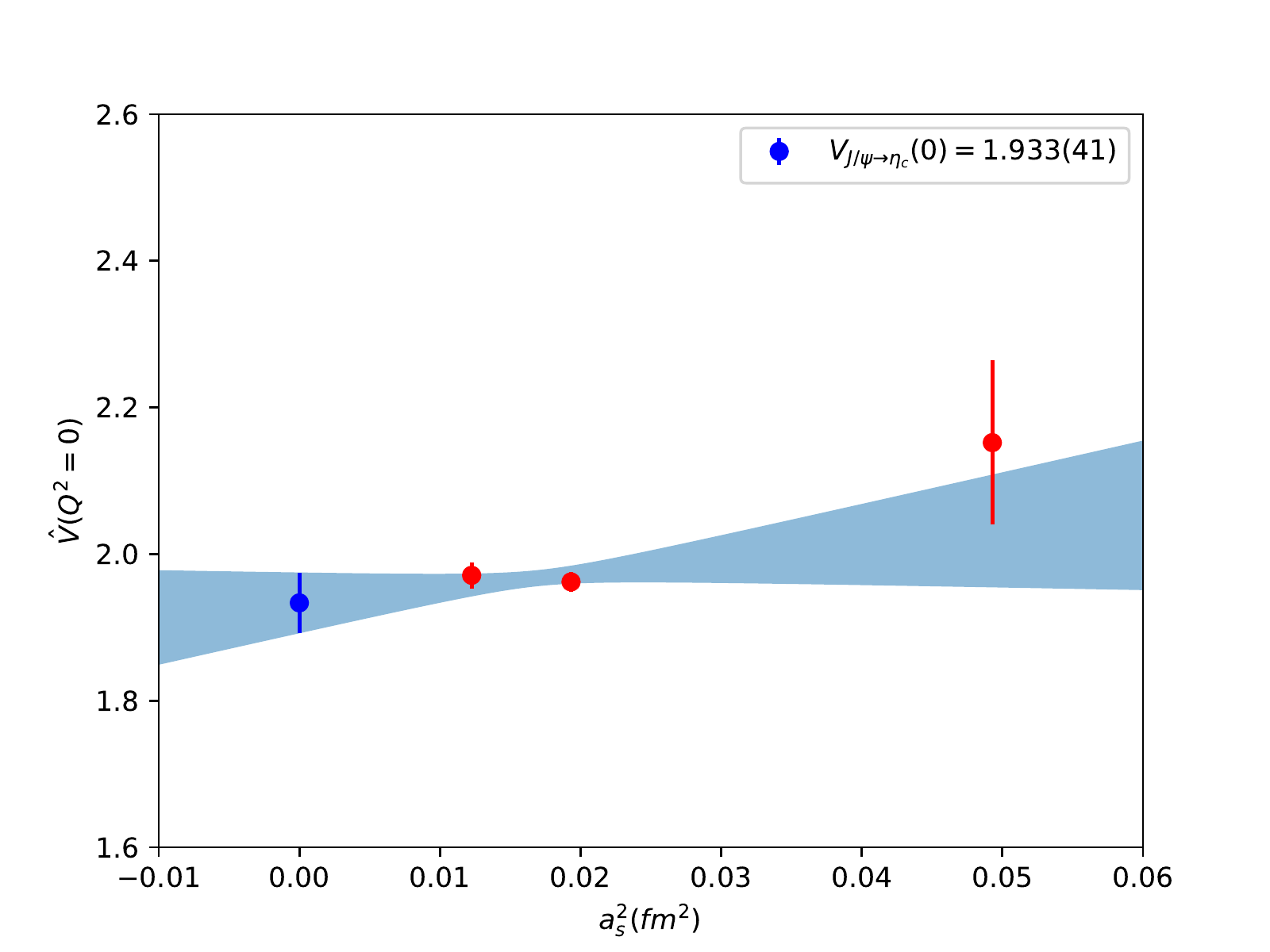} \caption{The continuum extrapolation of form factor $\hat{V}(Q^{2})$ of $J/\psi\to\gamma\eta_{c}$
on three lattices. The lattice spacing $a_{s}$ is in physical units,
$fm$. The blue band shows the continuum extrapolation linear in $a_s^2$.}
\label{fig:continuum_jpsitoetac}
\end{figure}

In order to obtain the on-shell form factor $\hat{V}(Q^{2}=0)$, we adopt
the following function form to do the extrapolation,
\begin{equation}
\hat{V}(Q^{2})=\hat{V}(0)e^{-\frac{Q^{2}}{16\beta^{2}}}\label{eq:jpsitoetacfitf}
\end{equation}
which is inspired by the simple quark model with the harmonic oscillator
wave functions of $\eta_{c}$ and $J/\psi$, as addressed in Ref.~\citep{Dudek2005a}.
The extrapolations are also illustrated in Fig.~\ref{fig:jpsitogammaetac}
by curves with error bands. It is interesting to see that this kind
of function form describes the data very well (On our coarsest $\beta=2.4$ lattice,
there is a clear deviation from the curve, which is tentatively attributed
to the relatively large discretization error of $Q^{2}$ calculated
on this lattice). The results after the extrapolation are listed in
Table~\ref{tab:jpsitogammaetac} and shown in Fig.\ \ref{fig:continuum_jpsitoetac}.
Since we have three lattices with different lattice spacings $a_{s}$,
we also perform a linear extrapolation
\begin{equation}
\hat{V}(0,a_{s})=\hat{V}(0)^{{\rm cont.}}+Aa_{s}^{2}
\end{equation}
to get the final result of $\hat{V}(0)^{{\rm cont.}}$,
\begin{equation}
\hat{V}(0)^{\rm cont.}=1.933(41),
\end{equation}
\textcolor{red}{}from which we give the prediction of the partial
width of the process $J/\psi\rightarrow\gamma\eta_{c}$,
\begin{eqnarray}
\Gamma_{J/\psi\to\gamma\eta_{c}} & = & \frac{\alpha|\vec{k}|^{3}}{(m_{J/\psi}+m_{\eta_{c}})^{2}}\frac{64}{27}|\hat{V}(0)|^{2}\nonumber \\
 & = & 2.47(11)\text{keV},
\end{eqnarray}
where the fine coupling constant takes the value at the charm quark
mass scale, $\alpha=\frac{1}{134}$, and $m_{J/\psi}$ and $m_{\eta_{c}}$
assume the experimental values.

%%%%%%%%%%%%%%%%%%%%%%%%%%%%%%%%%%%%%%%%%%%%%%%%%%%%%%%%%%%%%%%%%%%%%%%%%%%%%%%%%%%%%%%%%%%%%%%%
%%%%%%%%%%%%%%%%%%%%%%%%%%%%%%%%%%%%%%%%%%%%%%%%%%%%%%%%%%%%%%%%%%%%%%%%%%%%%%%%%%%%%%%%%%%%%%%%

We compare our result with those from previous lattice
QCD studies in Table~\ref{compare} where one can see that all the
results reach a consensus within errors. This assures us that, in
our study, the systematic uncertainties are not important in the charmonium
sector.

%%%%%%%%%%%%%%%%%%%%%%%%%%%%%%%%%%%%%%%%%%%%%%%%%%%%%%%%%%%%%%%%%%%%%%%%%%%%%%%%%%%%%%%%%%%%%%%%
\begin{table}
\caption{Comparison of $\hat{V}(0)$ with previous lattice results.\label{compare}}

\begin{ruledtabular}
\begin{tabular}{ll}
Lattice setup  & $\hat{V}(0)$ \tabularnewline
\hline
$N_{f}=2+1$ \citep{Davies2013}  & 1.90(7)(1) \tabularnewline
$N_{f}=2$ \citep{Becirevic2012}  & 1.92(3)(2) \tabularnewline
$N_{f}=2$~\citep{Chen2011}  & 2.01(2) \tabularnewline
Quenched~\citep{Dudek2005a}  & 1.85(4) \tabularnewline
Quenched (this work)  & 1.933(41)\tabularnewline
\end{tabular}
\end{ruledtabular}

\end{table}
%%%%%%%%%%%%%%%%%%%%%%%%%%%%%%%%%%%%%%%%%%%%%%%%%%%%%%%%%%%%%%%%%%%%%%%%%%%%%%%%%%%%%%%%%%%%%%%%

\subsection{The partial decay width of $J/\psi$ radiatively decaying into the pseudoscalar glueball}

We extend the similar study to the process of $J/\psi$ radiatively
decaying into the pseudoscalar glueball. It is known that the signals
of glueballs are always noisy, such that a large statistics is required.
On the other hand, an optimal interpolation operator, which couples
predominantly to the ground state $|G\rangle$ of the pseudoscalar
glueball, is mandatory for us to extract the desired matrix element
$\langle G|J^{\mu}_{{\rm em}}(0)|J/\psi\rangle$ reliably from the
related three-point functions in Eq.~(\ref{eq:threepf}). In doing
so, we adopt the strategy used in \citep{Morningstar1999,Chen2006}
to construct the optimal glueball operators, which is outlined as
follows. First, the last four of the ten Wilson loops in Fig.\ 3
of Ref.~\citep{Chen2006} are used as prototypes, and then six smearing
schemes (different combinations of the singlelike smearing and the
double-link smearing , as addressed in Ref.~\citep{Morningstar1999})
are applies to each of these prototype loops. Thus we obtain 24 different
Wilson loops as the basis operators. Since the lattice counterpart
of the quantum number $0^{-+}$ in the continuum is $A_{1}^{-+}$,
where $A_{1}$ is one of the five irreducible representations $A_{1},A_{2},E,T_{1},T_{2}$
of the spatial symmetry group $O$ of the cubic lattice, we apply
the 24 operations of $O$ group to each of the basis operators and
obtain 24 copies of it, whose proper linear combination gives the
representation of $A_{1}^{-+}$. Thereby we get 24 different operators
with the quantum number $A_{1}^{-+}$, which compose a operator set
$\{\phi_{\alpha},\alpha=1,2,...,24\}$. Based on this operator set,
we calculate the matrix of the correlation functions $\tilde{C}(t)=\{\tilde{C}_{\alpha\beta}\}$
with
\begin{equation}
\tilde{C}_{\alpha\beta}(t)=\frac{1}{T}\sum_{\tau}\langle\phi_{\alpha}(t+\tau)\phi_{\beta}(\tau)\rangle
\end{equation}
where we sum over $\tau$ to increase the statistics. Finally, by
solving the generalized eigenvalue problem
\begin{equation}
\tilde{C}(t_{D})V=\lambda(t_{D})\tilde{C}(0)V,
\end{equation}
we can get the eigenvector $V=\{v_{\alpha},\alpha=1,2,\ldots,24\}$
corresponding to the maximal eigenvalue $\lambda_{{\rm max}}(t_{D})\equiv e^{-m_{{\rm min}}(t_{D})t_{D}}$
($m_{{\rm min}}(t_{D})$ is close to the mass of the ground state),
from which we can obtain the optimal operator $\Phi(t)$ for the ground
state pseudoscalar glueball $|G\rangle$ by the combination $\Phi(t)=\sum_{\alpha}v_{\alpha}\phi_{a}$.
In this work, we set $t_{D}=1$.
\begin{figure}
\includegraphics[scale=0.45]{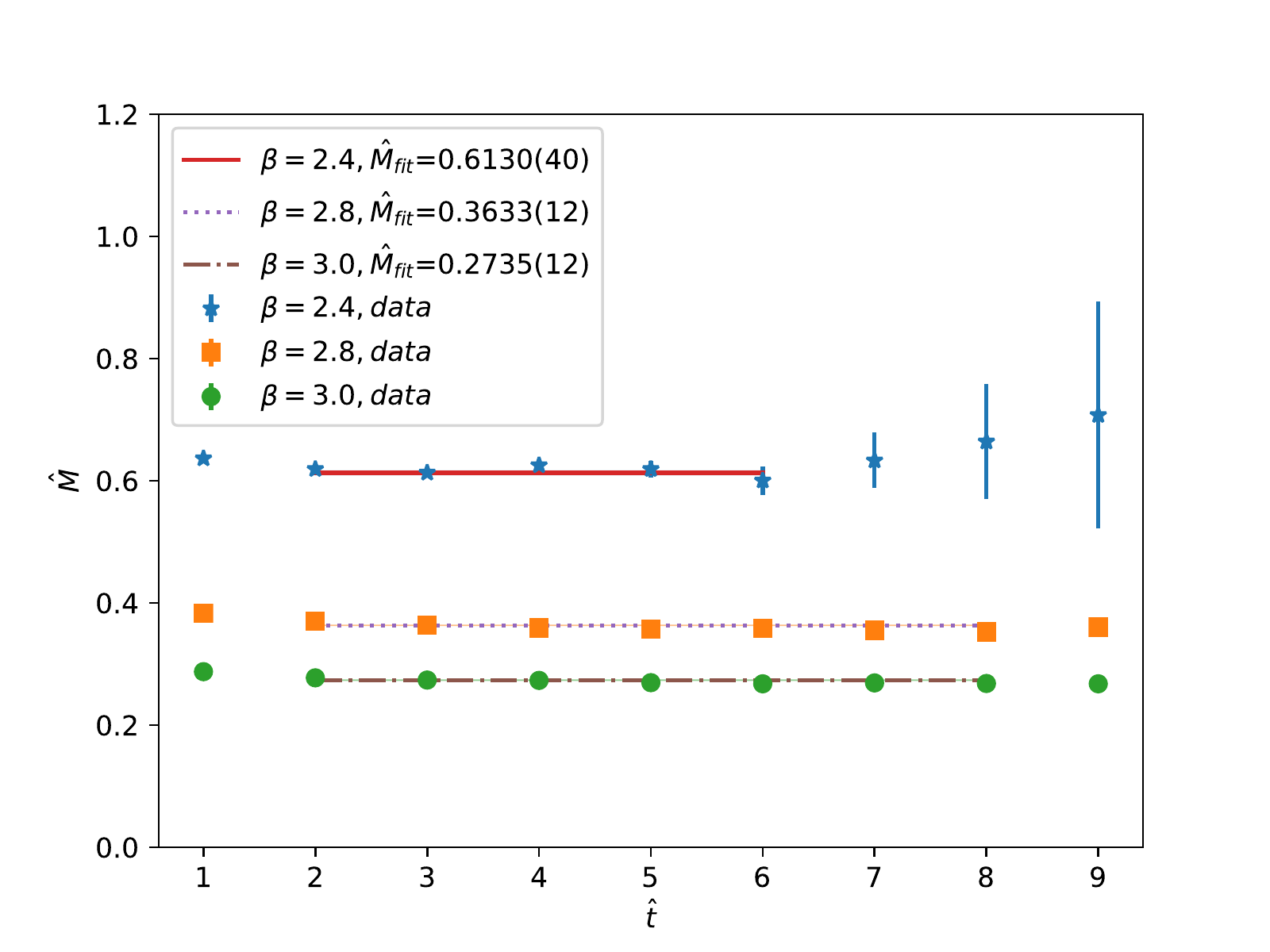} \caption{\label{fig:glueballeffectmass}The effect mass of pseudoscalar glueball
extracted from the two-point correlation function which constructed
by optimal glue operators on three lattices.}
\end{figure}
The correlation function of $\Phi(t)$ can be parametrized as
\begin{eqnarray}
C(t) & = & \frac{1}{T}\sum_{\tau}\langle\Phi(t+\tau)\Phi^{\dagger}(\tau)\rangle\nonumber \\
 & \simeq & \frac{Z_{G}^{2}}{2m_{G}V_{3}}e^{-m_{G}t}=We^{-m_{G}t},
\end{eqnarray}
where $m_{G}$ is the mass of the ground state, $V_{3}=L^{3}a_{s}^{3}$
is the spatial volume of the lattice, and $Z_{G}=|\langle0|\Phi(0)|G\rangle|$.
The effective mass of $C(t)$ is plotted in Fig.\  \ref{fig:glueballeffectmass},
where one can see that the plateau almost starts from the beginning
of time $t$. Obviously, there is still some contribution from higher
states which manifests by the slight increment of the effective mass
toward $t=0$. To check the extent of the higher state contamination,
we fit $C(t)$ through a single-exponential function and find that
the deviation of $W$ from one is at a level of few percents (note
that $C(t)$ is normalized as $C(0)=1$). This means that $C(t)$
is almost totally dominated by the contribution from the ground state
and therefore $W\simeq1$, or equivalently, $Z_{G}\simeq\sqrt{2m_{G}V_{3}}$
can be a good approximation. The mass of the ground state pseudoscalar
glueball on the three lattices are listed in Table~\ref{tab:jpsitoglueball}
. We obtain the mass of pseudoscalar glueball as $2.395(14)$ GeV
after continuum extrapolation. This value is lower than that in Ref.\ \citep{Chen2006},
but it is consistent within errors.

With the optimal operator $\Phi(t)$, the relevant three-point function
can be calculated and parametrized as
\begin{eqnarray}
&&\Gamma_{J/\psi  \to\gamma G_{0^{-+}}}^{(3),\mu i}(\vec{p}_{i},t_{i}=0;\vec{p}_{f},t_{f};\vecq,t)  \nonumber  \\ 
&& \quad  = \frac{1}{T}\sum\limits _{\tau,\vec{y}}e^{i\vec{q}\cdot\vec{y}}\langle\Phi(t_{f}+\tau) \times  J^{\mu}(\vec{y},t+\tau)\mathcal{O}^\dagger_{i}(\vec{0},\tau)\rangle\nonumber  \\
&& \quad \simeq  \sum\limits _{V,r}\frac{e^{-m_{G}(t_{f}-t)}e^{-E_{V}t}}{2m_{G}V_{3}2E_{V}}\langle \Omega |\Phi(0)|G\rangle\nonumber \\
&& \quad \times \langle G|J^{\mu}(0)|V(\vec{p}_{i},r)\rangle\langle V(\vec{p}_{i},r)|O_{i}^{\dagger}(0)| \Omega \rangle.
\end{eqnarray}

\begin{figure}[t]
\includegraphics[width=0.4\textwidth]{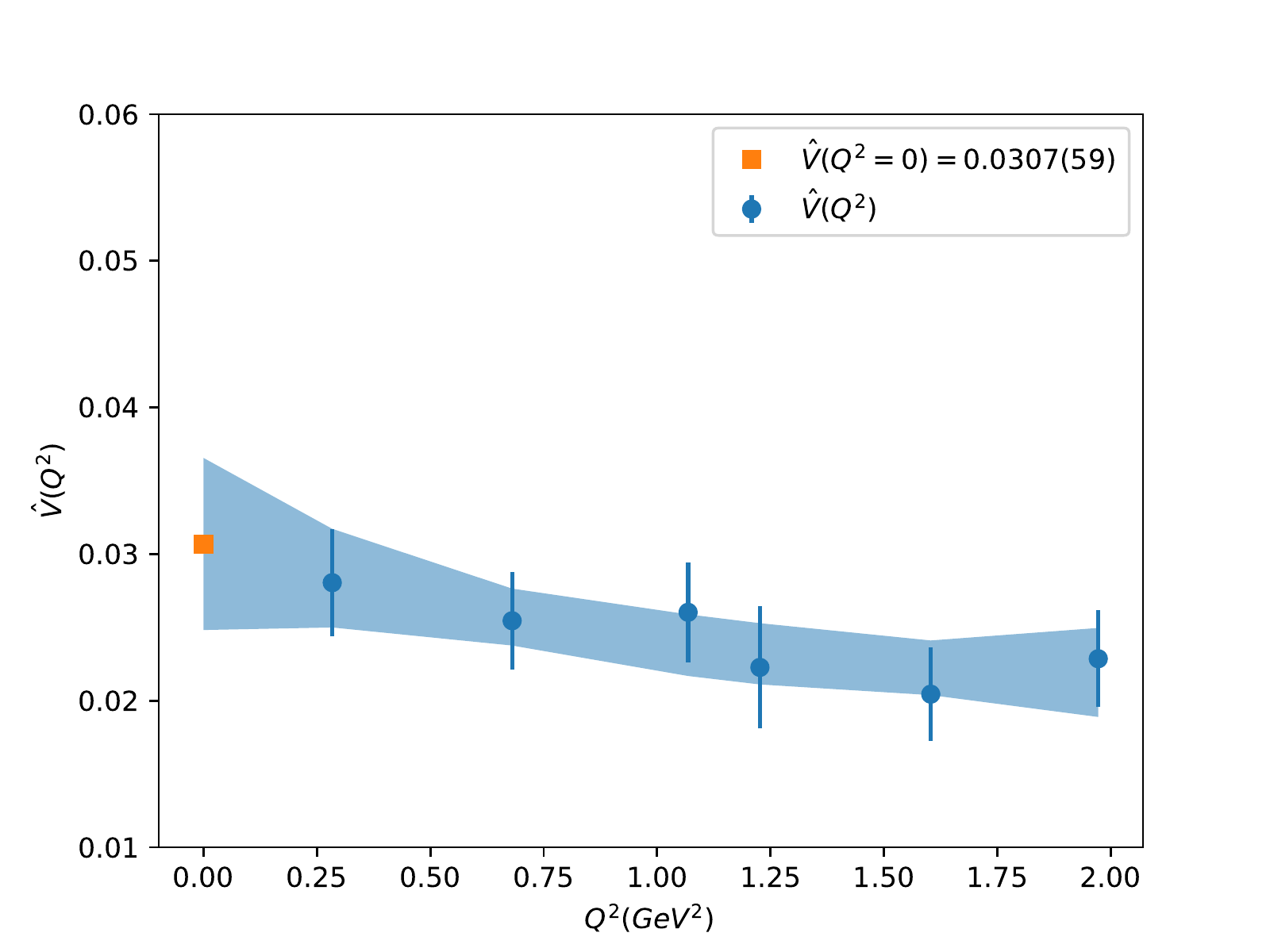}\\
 \includegraphics[width=0.4\textwidth]{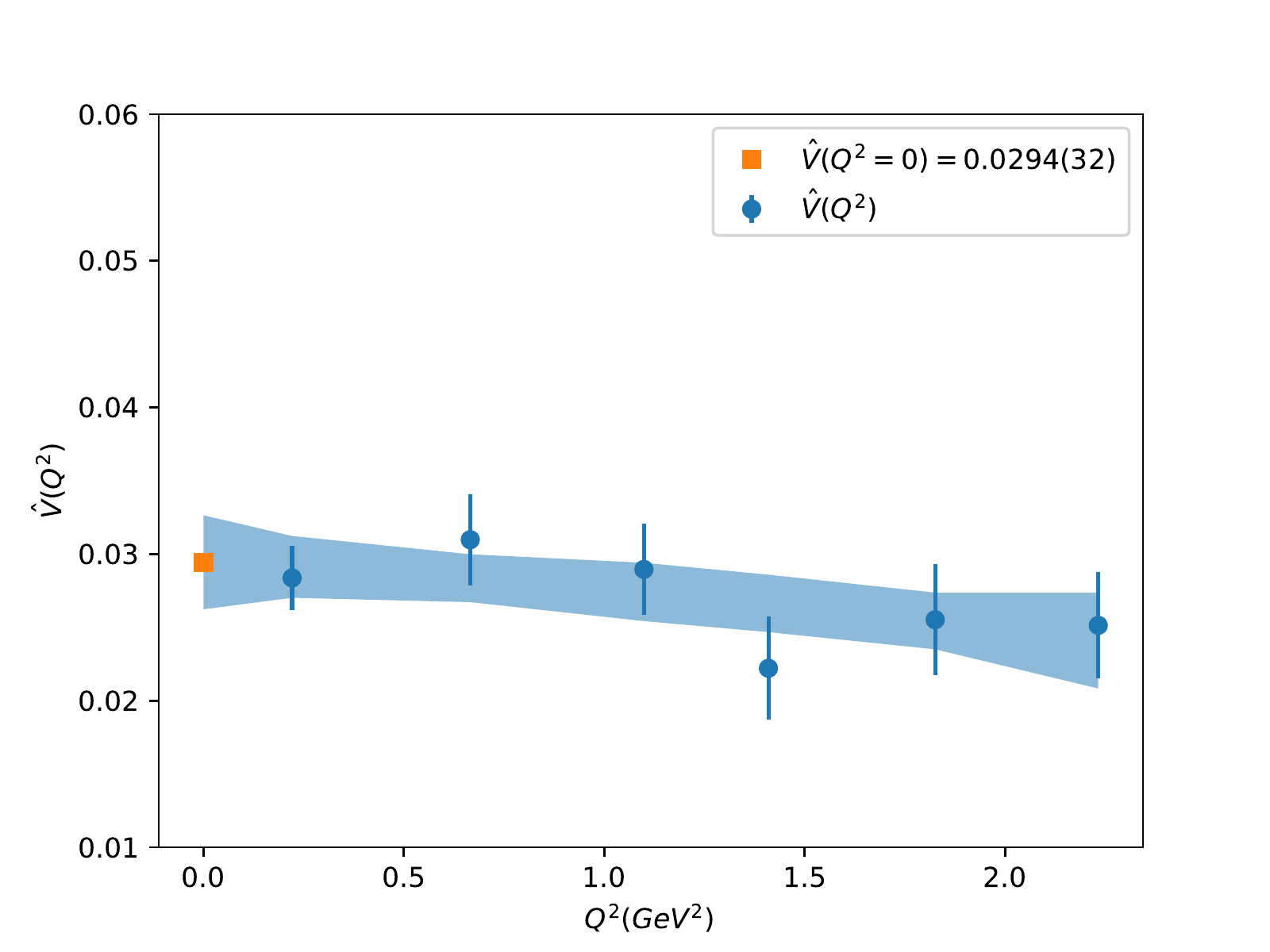}\\
 \includegraphics[width=0.4\textwidth]{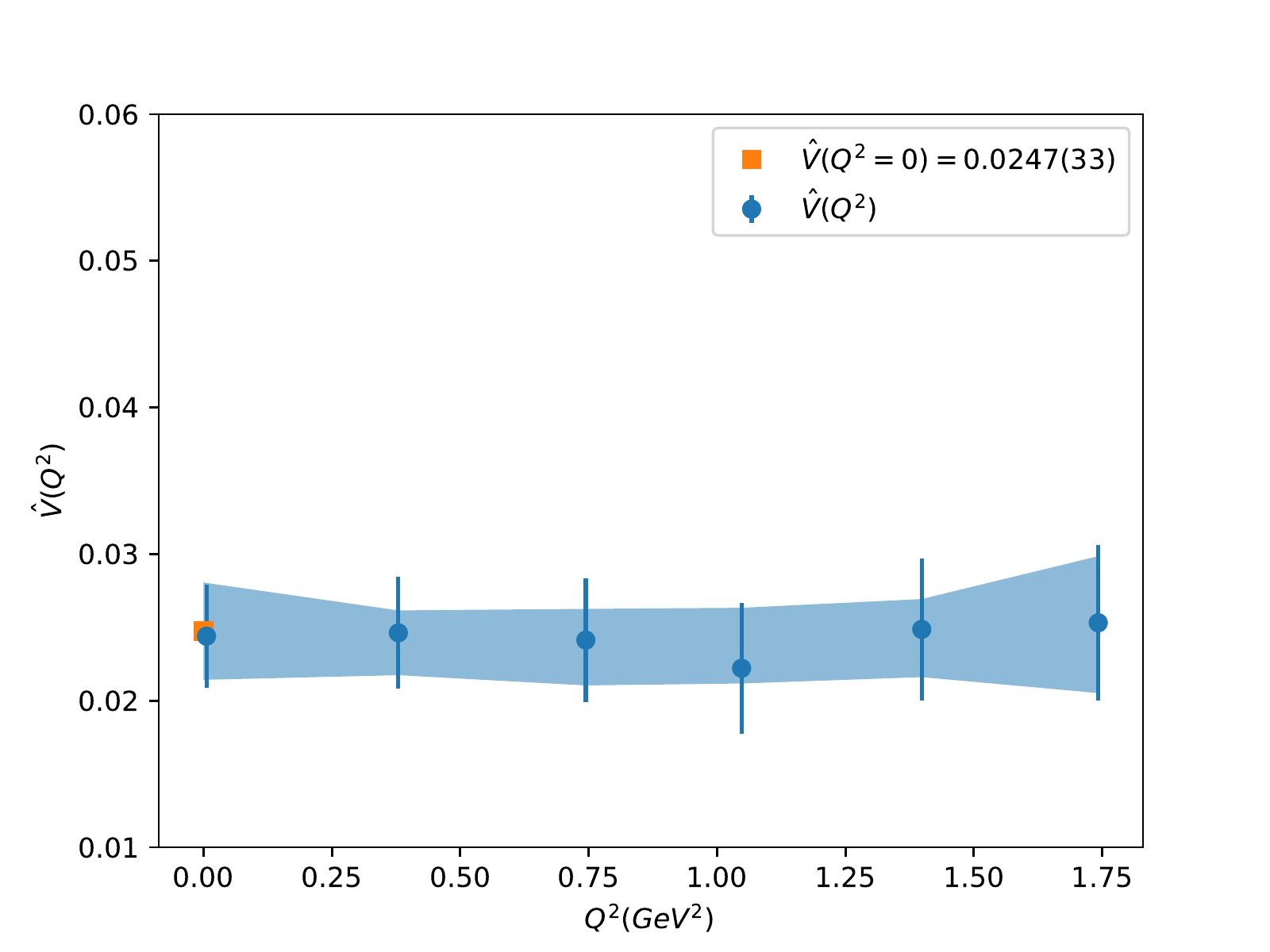}\\
 \caption{\label{fig:ffjpsitoglueball}The form factor $\hat{V}(Q^2)$ and the extrapolated
value $\hat{V}(Q^{2}=0)$ of $J/\psi\to\gamma G_{0^{-+}}$ where $Q^{2}=-(p_{J/\psi}-p_{G})^{2}$.
We use the function $\hat{V}(Q^{2})=\hat{V}(0)+aQ^{2}+bQ^{4}$ to perform the extrapolation.}
\end{figure}

Similar to the case of $J/\psi$ to $\eta_{c}\gamma$, we use this
formula %\begin{widetext}

\begin{eqnarray}
R^{\mu i}(\vecq,t_f,t) & = & \Gamma^{(3),\mu i}(\vec{q},t_{f},t)\frac{\sqrt{4V_{3}m_G E_{J/\psi}(\vec{q})}}{C(t_{f}-t)}\nonumber \\
 & \times & \sqrt{\frac{\Gamma_{J/\psi}^{(2)}(\vec{q},t_{f}-t)}{\Gamma_{J/\psi}^{(2)}(\vec{q},t)\Gamma_{J/\psi}^{(2)}(\vec{q},t_{f})}}
\end{eqnarray}
%\end{widetext}
to extract the matrix elements $\langle G|J^{\mu}|J/\psi\rangle$,
through which the contribution from excited states can be suppressed
to some extent. Practically, we fix the time interval of glueball operator and vector current operator
as one time slice which mean $t_{f}-t=1$, since that the optimal
glueball operator projects almost totally on the ground state pseudoscalar
glueball.  After the matrix elements $\langle G|J^{\mu}|J/\psi\rangle$ are derived, in analogy with 
the $J/\psi\to \gamma \eta_c$ case, we can 
obtain the form factors $\hat{V}(Q^2)$ at different $Q^2$, where $Q^{2}=-(p_{J/\psi}-p_{G})^{2}$ (note 
that $p_G=(m_G,\vec{0})$ since the pseudoscalar is at rest). The form factor $\hat{V}(Q^{2})$ derived on 
the three lattices are plotted in Fig.\  \ref{fig:ffjpsitoglueball} with respect to various $Q^2$. 

%%%%%%%%%%%%%%%%%%%%%%%%%%%%%%%%%%%%%%%%%%%%%%%%%%%%%%%%%%%%%%%%%%%%%%%%%%%%%%%%%%%%%%%%%%%%%
%%%%%%%%%%%%%%%%%%%%%%%%%%%%%%%%%%%%%%%%%%%%%%%%%%%%%%%%%%%%%%%%%%%%%%%%%%%%%%%%%%%%%%%%%%%%%%

\begin{figure}[t]
\includegraphics[scale=0.45]{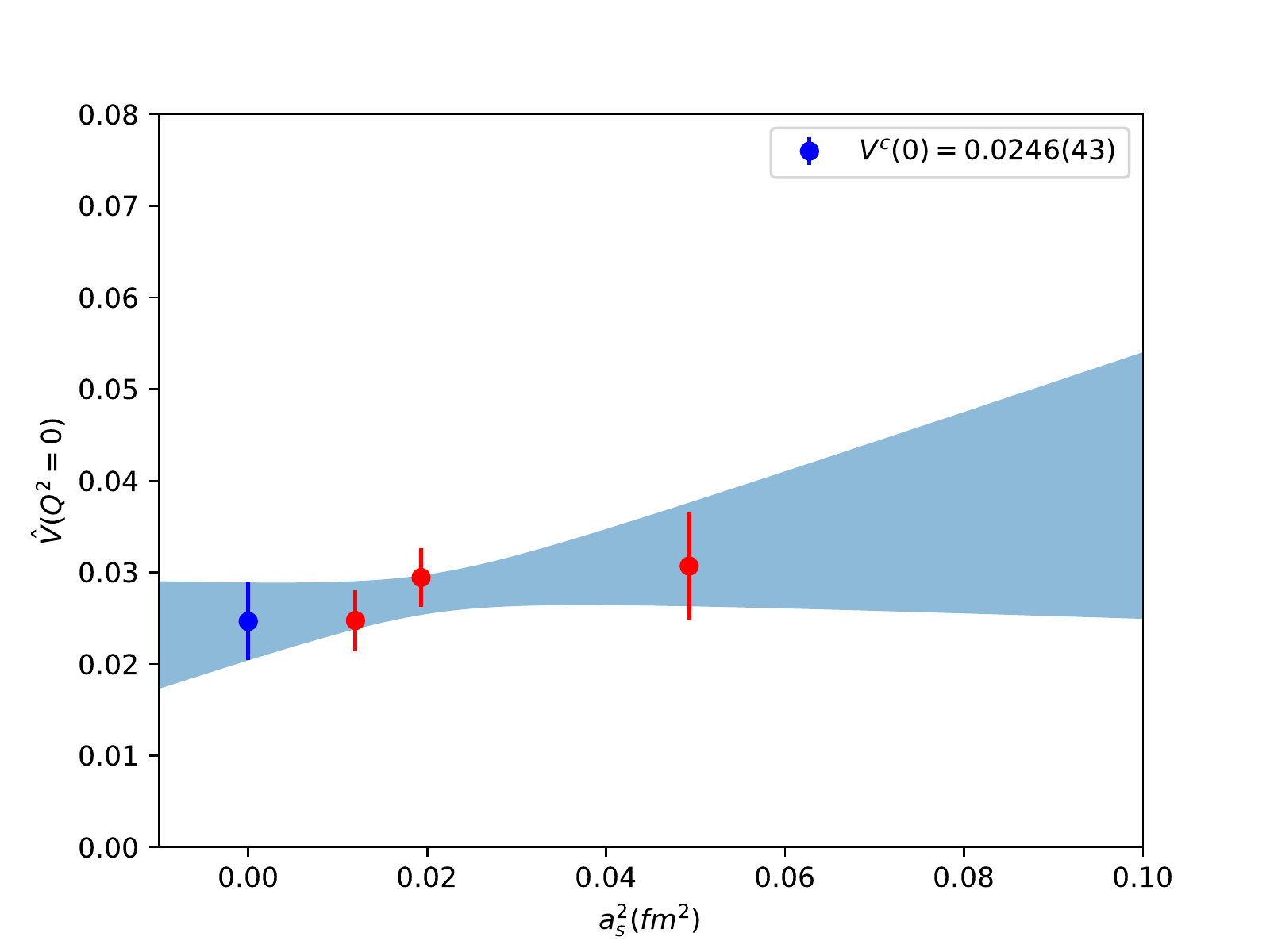} \caption{The continuum extrapolation of the form-factor $V(Q^{2}=0)$ of $J/\psi\to\gamma G_{ps}$.
The result show that the form-factor $V(0)$ is very weak dependence
to lattice spacing.}
\label{fig:continuum_jpsitoglueball}
\end{figure}

Because only the on-shell form factor, say, $\hat{V}(Q^2=0)$, enters into the formula of the transition width of $J/\psi\to\gamma G_{0^{-+}}$, 
\begin{equation}\label{eq:jpsitoglueballdecaywidth}
\Gamma(J/\psi\to\gamma G_{0^{-+}})=\frac{16}{27}\alpha\frac{|\vec{k}|^{3}}{(m_{G}+m_{J/\psi})^{2}}|\hat{V}(0)|^{2},
\end{equation}
we should perform an extrapolation of $\hat{V}(Q^2)$ from $Q^{2}\ne0$ to $Q^{2}=0$. However, in contrast 
to the case of $J/\psi\to \gamma \eta_c$ where the extrapolation function form of $\hat{V}(Q^2)$ (Eq.~(\ref{eq:jpsitoetacfitf})) can be 
inspired by the wave functions of $J/\psi$ and $\eta_c$ in the nonrelativistic quark model, we 
have no theoretical information for the $Q^2$ dependence of $\hat{V}(Q^2)$ in $J/\psi$ radiatively 
decaying into glueballs through the $\bar{c}c$ annihilation. Anyway, it is seen in Fig.~\ref{fig:ffjpsitoglueball} that $\hat{V}(Q^2)$ depends mildly on $Q^2$, therefore a polynomial 
fit in $Q^2$ can be safe here. The extrapolation formula is taken as 
\begin{equation}
\hat{V}(Q^{2})=\hat{V}(0)+aQ^{2}+bQ^{4}.\label{eq:extropfofjpsitoglueball}
\end{equation}
which can describe the data satisfactorily for all the three lattices, as shown in Fig.~\ref{fig:ffjpsitoglueball} with blue bands. The extrapolated $\hat{V}(0)$ on the three lattices 
are listed in Table~\ref{tab:jpsitoglueball}. In order to get the $\hat{V}(0)$ in the continuum 
limit, we also carry out a linear extrapolation in $a_s^2$ 
\[
\hat{V}(0)=\hat{V}(0)^{\rm cont.}+ca_s^2.
\]
Figure \ \ref{fig:continuum_jpsitoglueball} shows $\hat{V}(0)$ (red data points) at different lattice spacings and its continuum limit (blue point), where the blue band illustrates the linear extrapolation 
in $a_s^2$. Finally, the form factor $\hat{V}(0)$ in the continuum limit is determined to be
\begin{equation}
\hat{V}(0)^{\rm cont.}=0.0246(43),
\end{equation}
which gives the decay width of $J/\psi\to\gamma G_{0^{-+}}$ 
\begin{equation}
\Gamma(J/\psi\to\gamma G_{0^{-+}})=\begin{cases}
0.0215(74)\ {\rm keV} & m_{G}=2.395\ {\rm GeV}\\
0.0099(34)\ {\rm keV} & m_{G}=2.56\ {\rm GeV}
\end{cases}
\end{equation}
at different pseudoscalar masses according to Eq.~(\ref{eq:jpsitoglueballdecaywidth}). 
Consequently, the production fraction of the pseudoscalar glueball in the $J/\psi$ radiative decay is 
estimated to be 
\begin{equation}
Br(J/\psi\rightarrow\gamma G_{0^{-+}})=\begin{cases}
2.31(80)\times10^{-4} & m_{G}=2.395\ {\rm GeV}\\
1.07(37)\times10^{-4} & m_{G}=2.56\ {\rm GeV}
\end{cases}.
\end{equation}
%%%%%%%%%%%%%%%%%%%%%%%%%%%%%%%%%%%%%%%%%%%%%%%%%%%%%%%%%%%%%%%%%%%%%%%%%%%%%%%%%%%%%%%%%%%%
\begin{table}
\caption{\label{tab:jpsitoglueball}The mass of pseudoscalar glueball, $m_G$, and the
form-factor $\hat{V}(0)$ of $J/\psi\to\gamma G_{ps}$. The continuum limits of $m_G$ and $\hat{V}(0)$ 
are achieved by linear extrapolations in $a_s^2$.}

\begin{ruledtabular}
\begin{tabular}{ccc}
$\beta$ & $m_{G}$ (GeV)  & $\hat{V}(0)$ \tabularnewline
\hline
2.4  & 2.724(18)  & 0.0307(59) \tabularnewline
2.8  & 2.550(13)  & 0.0294(32) \tabularnewline
3.0  & 2.464(11)  & 0.0247(33) \tabularnewline
Continuum limit & 2.395(14) & 0.0246(43)\tabularnewline
  & 2.560(35)(120)\citep{Chen2006}  &  \tabularnewline
\end{tabular}
\end{ruledtabular}

\end{table}
%%%%%%%%%%%%%%%%%%%%%%%%%%%%%%%%%%%%%%%%%%%%%%%%%%%%%%%%%%%%%%%%%%%%%%%%%%%%%%%%%%%%%%%%%%%%
%%%%%%%%%%%%%%%%%%%%%%%%%%%%%%%%%%%%%%%%%%%%%%%%%%%%%%%%%%%%%%%%%%%%%%%%%%%%%%%%%%%%%%%%%%%%
%%%%%%%%%%%%%%%%%%%%%%%%%%%%%%%%%%%%%%%%%%%%%%%%%%%%%%%%%%%%%%%%%%%%%%%%%%%%%%%%%%%%%%%%%%%%

\section{Discussion}

Obviously, production fraction of the pure gauge pseudoscalar glueball
is quite small in the $J/\psi$ radiative decays, especially when
comparing with $3.8(9)\times10^{-3}$ of the pure gauge scalar glueball
and roughly 1\% of the tensor glueball. Furthermore, this value is
also much smaller than those of most known pseudoscalars. For example,
the branching fraction of $J/\psi\rightarrow\gamma\eta'$ is $5.13(17)\times10^{-3}$,
which is one orders of magnitude larger\textcolor{blue}{.} One of
the reasons is that $J/\psi$ radiatively decaying into a pseudoscalar
$X$ is through the $M1$ decay, such that the partial width is proportional
to $|\vec{k}|^{3}$ with $|\vec{k}|=(m_{J/\psi}^{2}-m_{X}^{2})/(2m_{J/\psi})$
being the magnitude of the decaying momentum of the final state photon.\textcolor{yellow}{{}
}Because the mass 2.4 GeV of the pseudoscalar glueball is close to
the mass of $J/\psi$, the partial width is suppressed by the kinematics.
In order to obtained a fair comparison, we would like to subtract
the phase space factor and introduce an effective coupling $g_{X}$
through the definition
\begin{equation}
\Gamma(J/\psi\rightarrow\gamma X)=\frac{1}{3}\alpha g_{X}^{2}\frac{|\vec{k}|^{3}}{m_{J/\psi}^{2}},
\end{equation}
where $1/3$ accounts for the spin average of the $J/\psi$, $\alpha$
is the fine coupling constant. Obviously, $g_{X}$ describes the coupling
of the gluons generated through $c\bar{c}$ annihilation to the pseudoscalar
meson $X$. Since in experiments only the branching fractions $Br(J/\psi\rightarrow\gamma X)$
can be measured directly, we express $g_{X}$ explicitly in terms
of these branching fractions along with the total width $\Gamma_{J/\psi}$
as
\begin{equation}
g_{X}=\left[\frac{24\Gamma_{J/\psi}}{\alpha}\frac{Br(J/\psi\rightarrow\gamma X)m_{J/\psi}^{5}}{(m_{J/\psi}^{2}-m_{X}^{2})^{3}}\right]^{\frac{1}{2}}.\label{effective-coupling}
\end{equation}
In practice, for $\eta$, $\eta'$, and $\eta(2225)$, we take the
branch fractions directly from the PDG data. For $\eta(1405)$ and
$\eta(1475)$, their production rates in the $J/\psi$ decays are
not differentiated from each other in PDG, so we also take them as
a whole and sum over the branching fractions of final states $\gamma K\bar{K}\pi$,
$\gamma\gamma\rho$, $\gamma\eta\pi^{+}\pi^{-}$, and $\gamma\rho^{0}\rho^{0}$.
For $\eta(1760)$ we add up the branching fractions of $\gamma\rho^{0}\rho^{0}$
and $\gamma\omega\omega$, for $X(1835)$ we add use the sum of the
branching ratios of the final states $\gamma+(\pi^{+}\pi^{-}\eta',p\bar{p},\eta K_{S}K_{S})$.
The branch fractions of $J/\psi $ radiative decay are listed in Table~\ref{tab:The-branch-fractions-PDG}
and the derived $g_{X}$'s are listed in Table~\ref{coupling}.

%%%%%%%%%%%%%%%%%%%%%%%%%%%%%%%%%%%%%%%%%%%%%%%%%%%%%%%%%%%%%%%%%%%%%%%%%%%%%%%%%%%%%%%%%%%%

\begin{table*}
\caption{\label{tab:The-branch-fractions-PDG}The branch fractions of $J/\psi$
radiative decay to pseudoscalar light mesons from the PDG data\citep{Tanabashi2018b}.}

\begin{ruledtabular}
\begin{tabular}{ccc}
pseudoscalar mesons & final states & branching ratios\tabularnewline
\hline
$\eta$ & $\gamma\eta$ & $(1.104\pm0.034)\times10^{-3}$\tabularnewline
\hline
$\eta^{\prime}$ & $\gamma\eta^{\prime}$ & $(5.13\pm0.17)\times10^{-3}$\tabularnewline
\hline
\multirow{4}{*}{$\eta(1405/1475)$} & $\gamma\eta(1405/1475)\rightarrow\gamma K\bar{K}\pi$ & $(2.8\pm0.6)\times10^{-3}$\tabularnewline
 & $\gamma\eta(1405/1475)\rightarrow\gamma\gamma\rho^{0}$ & $(7.8\pm2.0)\times10^{-5}$\tabularnewline
 & $\gamma\eta(1405/1475)\rightarrow\gamma\eta\pi^{+}\pi^{-}$ & $(3.0\pm0.5)\times10^{-4}$\tabularnewline
 & $\gamma\eta(1405/1475)\rightarrow\gamma\rho^{0}\rho^{0}$ & $(1.7\pm0.4)\times10^{-3}$\tabularnewline
\hline
\multirow{2}{*}{$\eta(1760)$} & $\gamma\eta(1760)\rightarrow\gamma\rho^{0}\rho^{0}$ & $(1.3\pm0.9)\times10^{-4}$\tabularnewline
 & $\gamma\eta(1760)\rightarrow\gamma\omega\omega$ & $(1.98\pm0.33)\times10^{-3}$\tabularnewline
\hline
\multirow{3}{*}{$X(1835)$} & $\gamma X(1835)\rightarrow\gamma\pi^{+}\pi^{-}\eta^{\prime}$ & $\left(2.77_{-0.40}^{+0.34}\right)\times10^{-4}$\tabularnewline
 & $\gamma X(1835)\rightarrow\gamma p\overline{p}$ & $\left(7.7_{-0.9}^{+1.5}\right)\times10^{-5}$\tabularnewline
 & $\gamma X(1835)\rightarrow\gamma K_{S}^{0}K_{S}^{0}\eta$ & $\left(3.3_{-1.3}^{+2.0}\right)\times10^{-5}$\tabularnewline
\hline
$\eta(2225)$ & $\gamma\eta(2225)$ & $\left(3.14_{-0.19}^{+0.50}\right)\times10^{-4}$\tabularnewline
\end{tabular}
\end{ruledtabular}

\end{table*}
\begin{table}
\caption{\label{coupling} The $g_{X}$ of flavor-singlet pseudoscalar mesons. }

\begin{ruledtabular}
\begin{tabular}{ll}
Pseudoscalar ($X$)  & $g_{X}$\tabularnewline
\hline
$\eta$  & 0.0108(2) \tabularnewline
$\eta^{\prime}$  & 0.0259(8)\tabularnewline
$\eta(1405/1475)$  & 0.0313(41)\tabularnewline
$\eta(1760)$  & 0.0255(25)\tabularnewline
$X(1835)$  & 0.0123(12)\tabularnewline
$\eta(2225)$  & 0.0167(17)\tabularnewline
 & \tabularnewline
$G_{0^{-+}}$  & 0.0126(22) \tabularnewline
\end{tabular}
\end{ruledtabular}

\end{table}
The striking observation is that the coupling $g_{X}$ for the pseudoscalar
glueball is comparable with or smaller than those of the known nonflavored
pseudoscalars (note that the smallness of $g_{\eta}$ is due to the
dominance of the flavor octet component of $\eta(547)$). This is
in sharp contrast to the usual expectation based on the naive $\alpha_{s}$-power
counting that the gluons in the $J/\psi$ radiative decay couple more
strongly to glueballs than $q\bar{q}$ mesons. Even though this naive
$\alpha_{s}$-power counting is not justified in the low energy QCD
regime, empirically there is an OZI rule observed in many hadronic
processes that the processes mediated through gluons, say, involving
the Feynman diagrams without continuous quark lines connecting the
initial and final states, tend to be strongly suppressed.

In this sense, the production of $\eta$ states in the $J/\psi$ radiative
decays seems to be OZI-violated. The QCD chiral anomaly may play an
important role in these processes. In QCD, the flavor singlet axial
vector current $j_{5}^{\mu}(x)=\sum\limits _{k=1}^{N_{f}}\bar{q}_{k}(x)\gamma_{5}\gamma^{\mu}q_{k}(x)$
is not conserved,
\begin{equation}
\partial_{\mu}j_{5}^{\mu}(x)=2N_{f}q(x)+\sum\limits _{k=1}^{N_{f}}2im_{k}\bar{q}_{k}(x)\gamma_{5}q_{k}(x),
\end{equation}
even in the chiral limit $m_{k}\to0$, where $N_{f}$ is the number
of the quark flavor, and the anomalous gluonic term $q(x)=\frac{g^{2}}{16\pi^{2}}trG_{\mu\nu}(x)\tilde{G}^{\mu\nu}(x)$
comes either from the regularization of the linearly divergent one-loop
diagrams of the vector-vector-axial vector current vertex (the triangle
diagram) in the perturbation theory or the chiral transformation noninvariance
of the fermion measure in the path integral formalism. As shown in
Ref. \citep{Gong2015,Yang:2019dha}, the matrix element related to
the chiral anomaly can be sizeable, and obviously violated the OZI
rule. In other words, the QCD $U_{A}(1)$ anomaly can enhance the
coupling of $\eta$ states to gluons, and this nonperturbative effect
results in the violation of the OZI rule when flavor singlet pseudoscalar
mesons are involved.

\textcolor{blue}{}

There is also evidence from lattice QCD study that the topological
charge density $q(x)$ couples to $\eta'$ strongly. A $N_{f}=2+1$
lattice simulation uses $q(x)$ to study $\eta'$ and observes a clear
contribution of $\eta'$ to the correlation function $\langle q(x)q(0)\rangle$~\citep{Fukaya2015}.
The authors get the result $m_{\eta'}=1019(119)$ MeV at the physical
pion mass consistent with the physical $\eta'$ mass. Similar lattice
studies have also been carried out in the $N_{f}=2$ case~\citep{Sun2017a,Dimopoulos2019}.
In Ref.~\citep{Sun2017a}, the authors get the mass of the isoscalar
pseudoscalar $\eta_{2}$ to be $m_{\eta_{2}}=890(38)$ MeV at $m_{\pi}=650$
MeV. In Ref.~\citep{Dimopoulos2019}, the authors use both the fermionic
operator $\bar{\psi}\gamma_{5}\psi$ and $q(x)$, and get compatible
results for $m_{\eta_{2}}$ on several gauge ensembles, whose chiral
extrapolation gives $m_{\eta_{2}}=772(18)$ MeV.

The discussion above helps to understand why the couplings of gluons
to $\eta$ states are not suppressed in comparison with the coupling
of the pseudoscalar glueball in the $J/\psi$ radiative decays. This
also implies the coupling $g_{X}$ cannot be used as a characteristics
for identifying the pseudoscalar glueball. Anyway, since the lattice
QCD studies predicts the pseudoscalar glueball mass is around 2.4-2.6
GeV, we may wish to check if there are any candidates in this mass
region. With the world largest $J/\psi$ event ensemble, the BESIII
Collaboration is performing a scrutinized partial wave analysis on
the radiative $J/\psi$ decay processes. In the process $J\psi\to\gamma\eta'\pi^{+}\pi^{-}$,
BESIII observes resonancelike structure $X(2120)$ and $X(2370)$
in the invariant mass spectrum of $\eta'\pi^{+}\pi^{-}$\citep{Ablikim2011,Ablikim2016b}
, however, their spin and parity have not been determined yet. $X(2370)$
is confirmed in the invariant mass spectra of $\eta'K\bar{K}$ by
BESIII~\citep{Prasad:2018fkb}, and its production fractions in the
processes $J/\psi\to\gamma X(2370)\to\gamma\eta'K^{+}K^{-}$ and $J/\psi\to\gamma X(2370)\to\gamma\eta'K_{S}K_{S}$
have been estimated to be $[1.86\pm0.39(sta.)\pm0.29(syss.)]\times10^{-5}$
and $[1.19\pm0.37(sta.)\pm0.18(syss.)]\times10^{-5}$, respectively.
In the partial wave analysis of the process $J/\psi\to\gamma\phi\phi$,
BESIII also observes a component $X(2500)$, whose preferred spin-parity
assignment is pseudoscalar~\citep{Ablikim2016a}. Its production
fraction in the process $J/\psi\to\gamma X(2500)\to\gamma\phi\phi$
is determined to be $[1.7\pm0.2_{-0.8}^{+0.2}]\times10^{-5}$. Whether
they are the same object or not, $X(2370)$ and $X(2500)$ reside
in the pseudoscalar glueball mass region predicted by lattice QCD,
whose production rates are compatible with that of the pseudoscalar
glueball in this work. Recently, BESIII has finished the data collection
of 10 billion $J/\psi$ events, hopefully the properties of these
states can be determined more precisely in the near future.

\section{SUMMARY}

We carry out the first lattice calculation of the production rate
of the pseudoscalar glueball in $J/\psi$ radiative decays in the
quenched approximation. We generate gauge configurations on three
anisotropic lattices with different lattice spacings, which facilitate
us to perform the continuum limit extrapolation. In order to calibrate
some of the systematic uncertainties, we first calculate partial decay
width of $J/\psi\to\gamma\eta_{c}$. The related $M_{1}$ transition
form factor is determined to be $\hat{V}(0)=1.933(41)$, which gives
the partial width $\Gamma(J/\psi\to\gamma\eta_{c})=2.47(11)$ keV.
These results are consistent with the results of previous lattice
calculations.

By applying the variational method to a large operator set, we obtain
an optimal operator which couples predominantly to the ground state
pseudoscalar glueball $G$. In this work, $m_{G}$ is determined to
be 2.395(14) GeV, and the on-shell form factor of $J/\psi\to\gamma G$
is derived as $\hat{V}(0)=0.0246(43)$, in the continuum limit, from
which we obtain the following partial decay width and the production
rate
\begin{align}
\Gamma(J/\psi\to\gamma G_{0^{-+}}) & =0.0215(74)\ keV\nonumber \\
Br(J/\psi\to\gamma G_{0^{-+}}) & =2.31(80)\times10^{-4}
\end{align}
We introduce an effective coupling $g_{X}$ to describe the interaction
between the pseudoscalar $X$ and the gluonic intermediate states
in the processes $J/\psi\to\gamma X$, as defined in Eq.~(\ref{effective-coupling}).
It is interesting to see that all the $g_{X}$'s are comparable for
the pseudoscalar glueball and the nonflavored $q\bar{q}$ pseudoscalars
($\eta$ states). We tentatively attribute the large production rates
of the $\eta$ states to the QCD $U_{A}(1)$ anomaly which is totally
a nonperturbative effect.

Even though this study is performed in the quenched approximation
and the uncertainty in the presence of dynamical quarks is not controlled,
we hope our result can provide useful theoretical information for
experiments to unravel the properties of the possible pseudoscalar
glueball.

\section*{Acknowledgements}
L.C. thanks J. Liang for useful discussions. This work is supported
by the National Key Research and Development Program of China (No.2017YFB0203202).
The numerical calculations are carried out on Tianhe-1A at the National
Supercomputer Center (NSCC) in Tianjin and the GPU cluster at Hunan
Normal University. We acknowledge the support of the National Science
Foundation of China (NSFC) under Grants No. 11575196, No. 11405053,
No. 11335001, No. 11621131001 (CRC 110 by DFG and NSFC), No. U1832173, No. 11705055. Y.C. is
also supported by the CAS Center for Excellence in Particle Physics
(CCEPP) and National Basic Research Program of China (973 Program)
under code number 2015CB856700. Y.Y. is supported by the CAS Pioneer
Hundred Talents Program. Our matrix inversion code is based on QUDA
libraries\citep{Clark2010a}and the fitting code is based on lsqfit\citep{Lepage:2001ym}
.

\end{document}